\documentclass[aps,prl,superscriptaddress,twocolumn,longbibliography]{revtex4-1}
\pdfoutput=1

\setcitestyle{super}
\usepackage[T1]{fontenc}
\usepackage{comment}
\usepackage{graphicx}
\usepackage{bm}
\usepackage{amsfonts,mathtools,amssymb,stmaryrd}
\usepackage{color}
\usepackage[breaklinks]{hyperref}
\usepackage{epstopdf, epsfig}
\usepackage{natbib}
\usepackage{grffile}
\usepackage{amsthm}

\newcommand{\bX}{{\bf X}}
\newcommand{\bx}{{\bf x}}
\newcommand{\bR}{{\bf R}}
\newcommand{\br}{{\bf r}}
\newcommand{\be}{\begin{equation}}
\newcommand{\ee}{\end{equation}}
\newcommand{\bsp}{\begin{split}}
\newcommand{\esp}{\end{split}}

\newcommand{\bi}{\begin{itemize}}
\newcommand{\ei}{\end{itemize}}

\newcommand{\av}[1]{\left\langle#1\right\rangle}
\newcommand{\avY}[1]{\mathbb E\left(#1\right)}

\newcommand{\inphyniaddress}{Institut de Physique de Nice, Universit\'e C\^ote d’Azur CNRS - UMR 7010, 17 rue Julien Lauprêtre, 06200 Nice, France}
\newcommand{\impaaddress}{Instituto de Matem\'atica Pura e Aplicada -- IMPA, 22460-320 Rio de Janeiro, Brazil}
\newcommand{\ceaaddress}{SPEC, CEA, CNRS, Université Paris-Saclay, CEA Saclay, Gif-sur-Yvette, France}

\newcommand{\corr}{\color{black}}
\newcommand{\rroc}{\color{black}}

\graphicspath{{Figs/}}

\begin{document}

\title{Spontaneous stochasticity in the presence of intermittency}
\author{André Luís Peixoto Considera}
\affiliation{\impaaddress}
\affiliation{\ceaaddress}
\author{Simon Thalabard}
 \affiliation{\inphyniaddress}
\date{\today}
\begin{abstract}
Spontaneous stochasticity is  a modern paradigm for turbulent transport at infinite  Reynolds numbers. It suggests that  tracer particles advected by rough turbulent flows and subject to additional  thermal noise,  remain non-deterministic   in the limit where the random input, namely  the thermal noise,  vanishes. Here, we investigate the fate of spontaneous stochasticity in the presence of spatial intermittency, with multifractal scaling of the lognormal type,  as usually encountered in turbulence studies.  In principle, multifractality enhances the underlying roughness, and \corr should also favor  the  spontaneous stochasticity\rroc. This letter  exhibits a case with a \corr less \rroc intuitive  interplay between  spontaneous stochasticity and spatial intermittency. We specifically address Lagrangian transport in  unidimensional multifractal random flows, obtained by decorating  rough Markovian monofractal  Gaussian  fields with frozen-in-time Gaussian multiplicative chaos.  Combining  systematic Monte-Carlo simulations  and formal stochastic calculations, we  evidence a  transition between spontaneously stochastic   and  deterministic behaviors when increasing the level of intermittency. While its key ingredient in the Gaussian setting,  \corr  roughness here suprisingly \rroc conspires against the spontaneous stochasticity of trajectories.
\end{abstract}

\maketitle
\paragraph{Introduction.}
When transported by a sufficiently turbulent flow, puffs of fluid particles are known to undergo a phase of algebraic inflation  $R\sim t^{3/2}$, independent from \corr their initial size \rroc and now known as Richardson diffusion \cite{richardson1926atmospheric,jullien1999richardson,boffetta2002statistics, boffetta2002relative,bitane2012time, thalabard2014turbulent, bourgoin2015turbulent, buaria2015characteristics, tan2022universality,  buaria2023comment}. Beyond the  law in itself,  Richardson's seminal contribution is the intuition that  turbulent transport requires some probabilistic
modeling: 
The modern interpretation \corr uses \rroc the phenomenon of spontaneous stochasticity\cite{lorenz69predictability,mailybaev2016spontaneously,mailybaev2016spontaneous,thalabard2020butterfly,mailybaev2022spontaneously,mailybaev2022spontaneous}, which 
involves tracers as  fluid particles advected by the fluid and subject to additional thermal noise of amplitude $\kappa$ \cite{gawedzki2008soluble}:
 In the vanishing viscosity limit, the multi-scale nature of turbulent flows amplifies thermal noise in such a drastic fashion,  that initially coinciding particles may separate in finite time although \corr their dynamics \rroc formally solve the same initial value problem \cite{e2000generalized,kupiainen2003nondeterministic,chaves2003lagrangian}, hereby suggesting intrinsic nature for the underlying randomness. \\
To date, the scenario of spontaneous stochasticity for Lagrangian separation is fully substantiated within the theory of Kraichnan flows. Kraichnan flows are minimal random \corr ersatzes \rroc of homogeneous isotropic turbulent fields \cite{falkovich2001particles,le2002integration,le2004flows,kupiainen2003nondeterministic,gawedzki2001turbulent,gawedzki2008soluble}; They are defined as    white-in-time  Gaussian random fields, whose    spatial statistics are centered and  prescribed  by two-point  correlation \corr functions \rroc with  algebraic decay of the kind
\be
  C_\eta^{(\xi)}(r)  
	 = 1-|r|^{\xi}  \text{ for $\eta \le |r| \ll 1$},
\label{eq:correlation}
\ee
and vanishing at large scales $\gg 1$. $\eta$ is a scale under which the flow is smooth, analogous to so-called Kolmogorov scale: The scales $\eta \le |r| \ll 1$ define  the so-called inertial range  in turbulence theory. The Hurst parameter $\xi \in ]0;2[$  prescribes the roughness of the field, through inertial-range  scaling $\av{(v(x+r)-v(x))^2} \sim r^\xi$. In the limit $\eta \to 0$,  this means that  the lesser $\xi$, the rougher $v$. 
In this stochastic setting,   spontaneous stochasticity  essentially means that some random
time accounting for  the large-scale $O(1)$ dispersion  of a puff of tracers with initial size $O( \eta)$  has probability 1 to be finite
 in  limits where $\eta,\kappa$ jointly vanish. The limit describes  puffs initially coalescing to a point in  prescribed (quenched) space-time velocity realizations \cite{boffetta2002relative, chaves2003lagrangian}.
For instance, explicitly considering the relative separation $\bR(t,\br_0) := \bX_2(t,\bx_0+\br_0) -\bX_1(t,\bx_0)$  between two tracers initiated  at  $\bf x_0, \bf x_0+\br_0$, a  natural separation time  is  
\be
	\tau_1(\eta,\kappa ) := \inf_{\|\br_0\|=\eta} \left\lbrace t | \|\bR(t,\br_0)\| \ge 1 \right\rbrace.
	\label{eq:exit}
\ee
from which  we  interpret spontaneous stochasticity as the property
\be
	\mathbb P \left[ \tau_1 < \infty \right] \to 1 \text{ as $\eta,\kappa \to 0$}.
	\label{eq:ss}
\ee
Even at this essential level, the presence or the absence of spontaneous stochasticity in Kraichnan flows depends on a subtle interplay between four parameters: roughness,  \corr compressibility\rroc, space-dimension,  reflection rules for colliding trajectories.
To highlight the effect of roughness, we focus on the unidimensional space, hence prescribing unit compressibility, with a thermal noise  $\kappa = \eta $ ensuring that  colliding trajectories reflect upon collision in the limit $\eta \to 0$. The only relevant parameter is then the roughness exponent $\xi$: Spontaneously stochastic property ~\eqref{eq:ss}  holds if and only if   $\xi<1$. For $\xi \ge 1$, particles wind up sticking together hence producing apparent deterministic behavior \cite{gawedzki2000phase,gawedzki2008soluble,supplemental}: 
 In short, Kraichan flows suggest the mantra ``The rougher, the more spontaneously  stochastic''.
In this letter, we show that \corr this \rroc mantra cannot be repeated  in the presence of multifractality, a feature which we later also refer to as   spatial intermittency.\\

\begin{figure}[h]
\includegraphics[width=0.485\columnwidth]{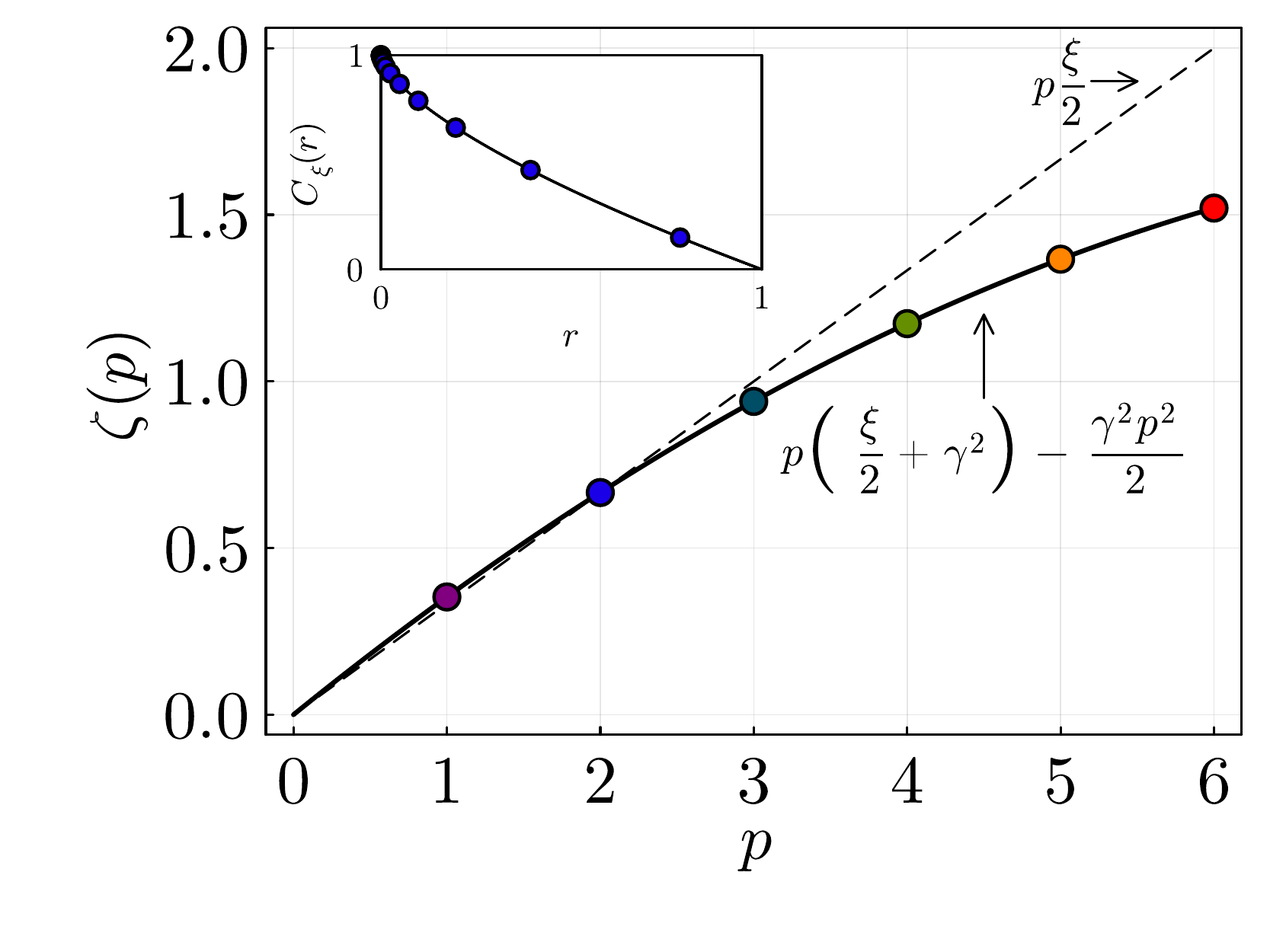}
\includegraphics[width=0.50\columnwidth,trim=0.02cm 0cm 0.0cm 0cm , clip]{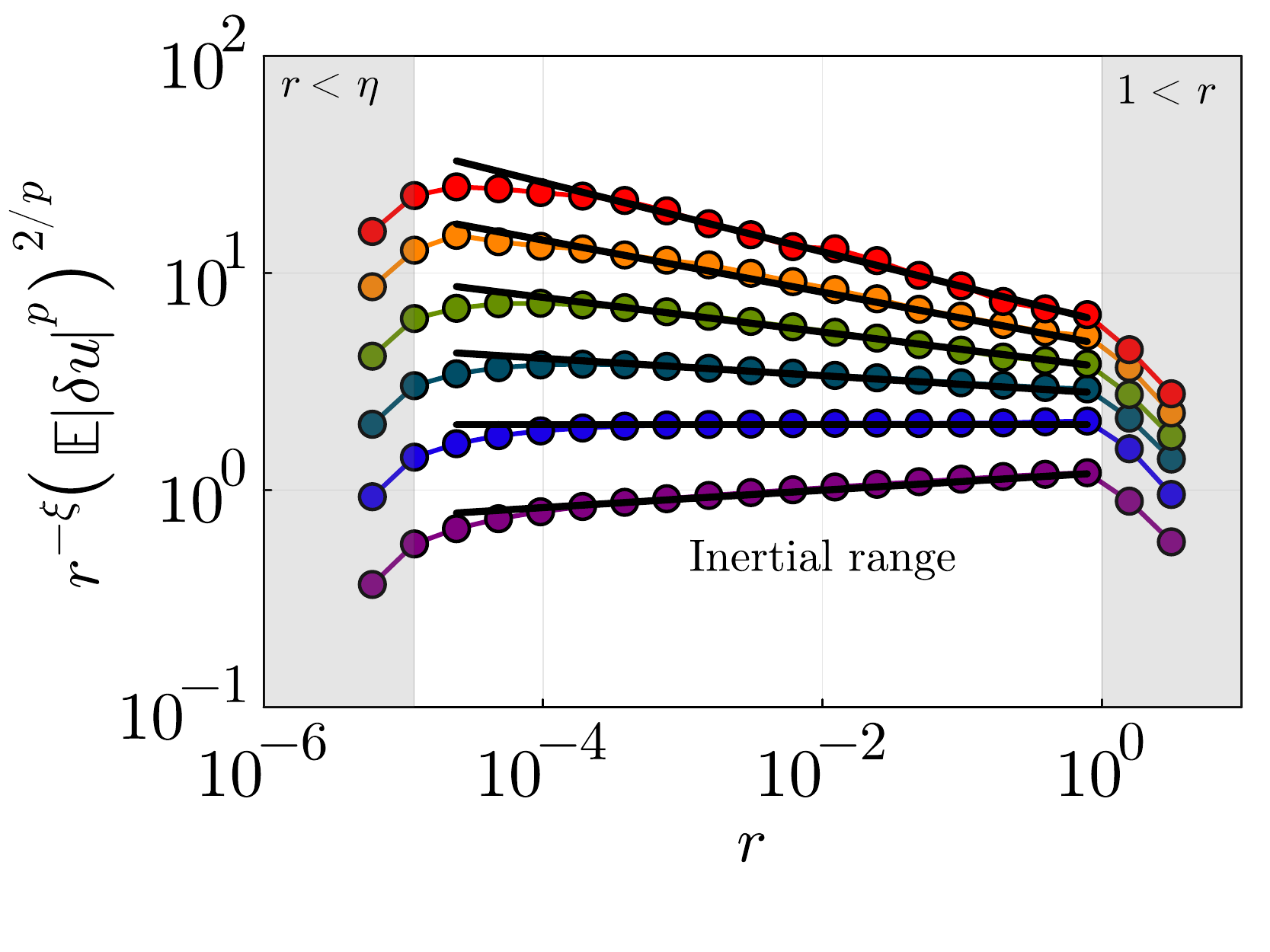}\\
\includegraphics[width=0.49\columnwidth,trim=0.25cm 0cm 0.25cm 0cm , clip]{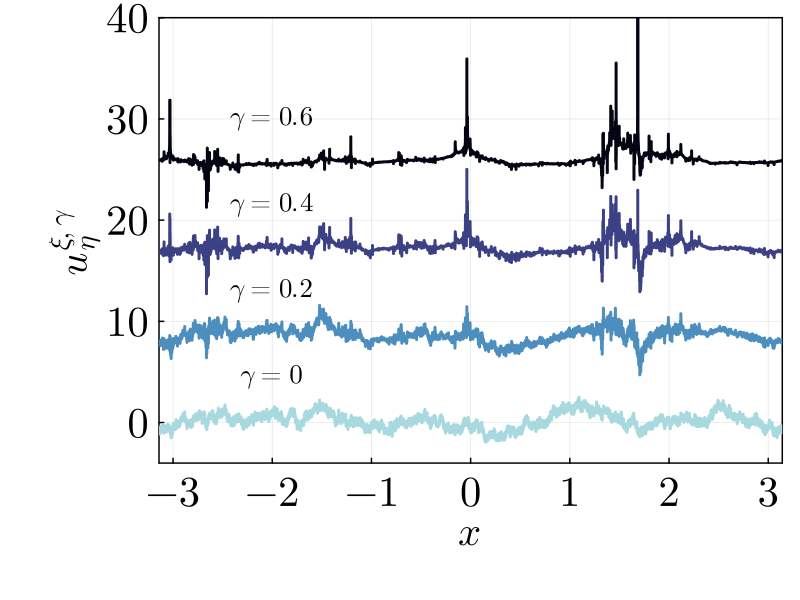}
\includegraphics[width=0.49\columnwidth]{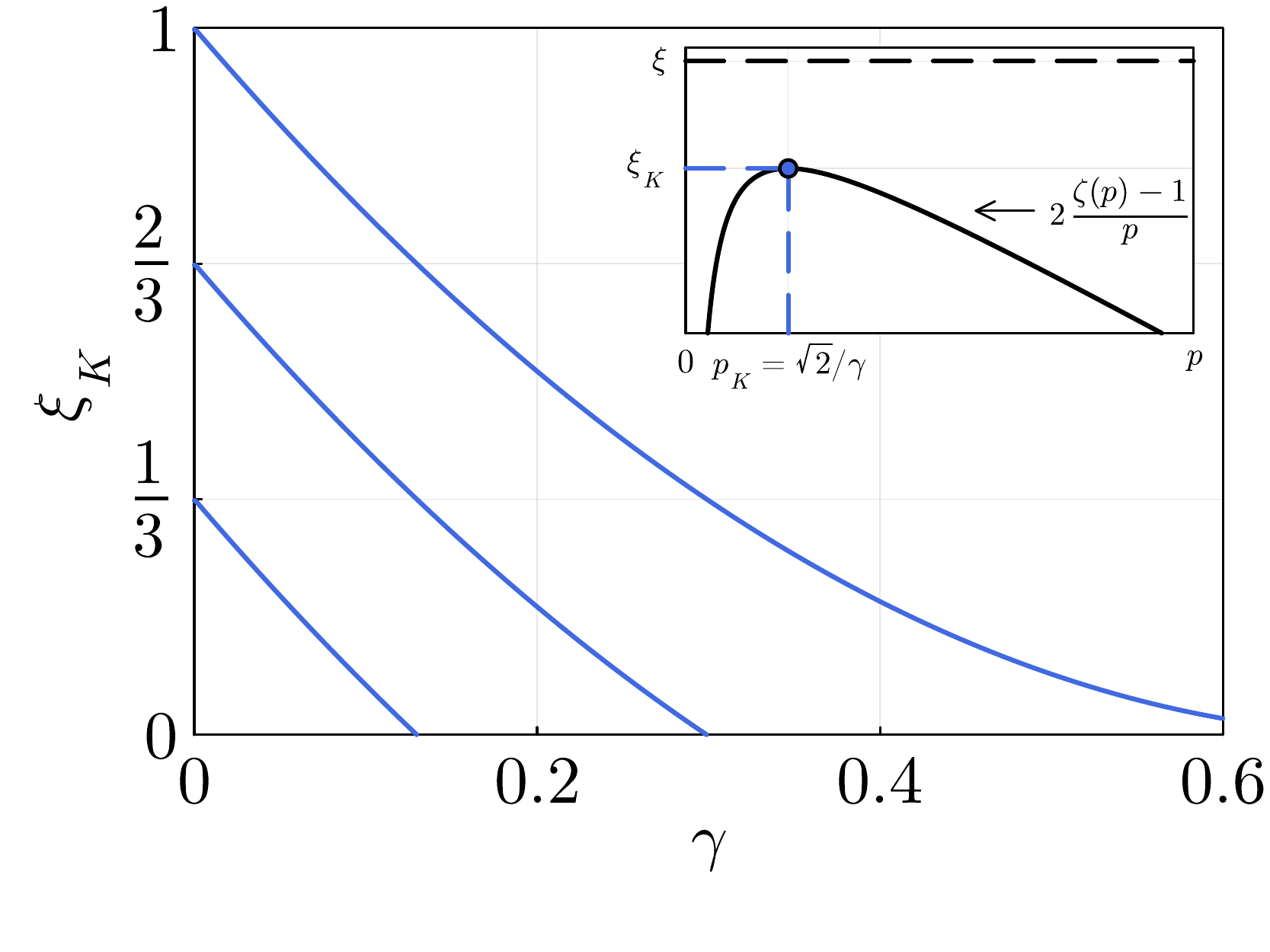}
\setlength{\unitlength}{\columnwidth}
\begin{picture}(1,0)(0,0)
\put(0.11,0.53){(a)}
\put(0.6,0.53){(b)}
\put(0.11,0.14){(c)}
\put(0.59,0.14){(d)}
\end{picture}
\caption{ (a)  Multifractal prediction  given by Eq.~\eqref{eq:expo}  for $\gamma=0.2$.  Inset shows the correlation function with  the power-law decay \eqref{eq:correlation} for $\xi=\zeta(2)$.
(b) Compensated structure functions $r^{-\xi}\av{|\delta u|^p }^{2/p} \propto r^{2\zeta(p)/p-\xi}$  for integer orders $1 \le p \le 6$ (from bottom to top). Points indicate  Monte-Carlo averaging over $512$ samples using $N=2^{20}$ points and smoothing scale $\eta=4\pi/N$. The black shaded lines indicate the multifractal prediction. (c) Random realizations of the spatial part of \eqref{eq:MarkovGMC}
for $\xi=2/3$ and various $ \gamma$ . (d) Effective roughess $\xi_K$ as a function of $\gamma$ for various $\xi$. Inset illustrates  Definition ~\eqref{eq:effectiveroughness}. }
\label{fig:1}
\end{figure}
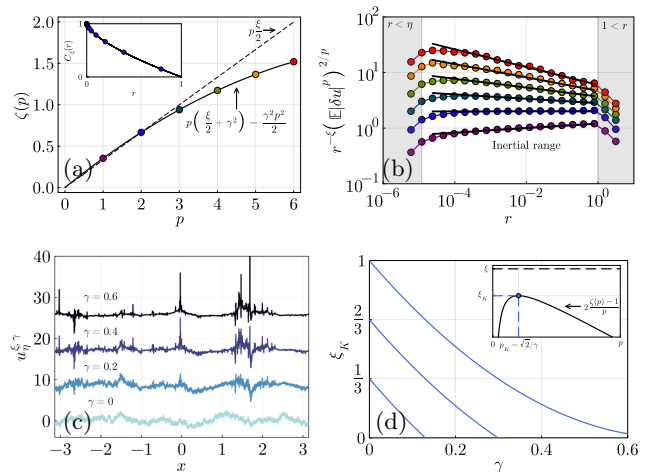

\paragraph{1D multifractal Kraichnan flows.}
We propose a multifractal unidimensional generalization of the Kraichnan model, which
 prescribes  the motion  of $N$ tracers particles $(X_i,i \in [|1;N|])$ in terms of the advection-diffusion
\be
	dX_{i} = u_{\eta}^{\xi,\gamma}\left( X_i(t), dt \right) +\sqrt{ 2\kappa}\; B_i(dt),
	\label{eq:dynamics}
\ee
where the $B_i$'s are independent Brownian motions and we set $\kappa = \eta$ for the thermal noise amplitude:   This scaling 
ensures that close-by tracers  separated by at most $\eta$ diffuse away from each other. The smoothing scale $\eta$ will ultimately be \corr taken \rroc to $0$.
 The velocity  $u_{\eta}^{\xi,\gamma}$ models turbulent advection in a  rough multifractal field,
prescribed by the Hurst exponent $\xi \in ]0;2[$ and the intermittency parameter $\gamma$.   We use the 1D Markovian version of the spatio-temporal fields constructed by Chevillard \& Reneuve \cite{reneuve2020flow} 
\be
	\begin{split}
	& u^{\xi,\gamma}_\eta(x,dt)  = \dfrac{1}{Z_\eta}\int_{\mathbb R} L^{(\xi)}_{\eta}(x-y)\,e^{\gamma Y(y)}  W_1(dy,dt),\\
	& \text{for}\;
	Y(y) := \int_{\mathbb R} L_{\eta}^{(0)}(y-z) W_2(dz),\;\; Z:= e^{ \gamma^2 \mathbb E\left({Y^2}\right)},
	\end{split}
	\label{eq:MarkovGMC}
\ee
in terms of the mutually independent  (1+1)  dimensional Wiener process $W_1$ and Brownian motion $W_2$,  also independent from the $B_i$'s.  
 The kernels $L^{(\xi)}_{\eta}$ prescribe the Hurst exponent of the velocity field when $\gamma=0$; They are here defined as convolution square roots of the correlation function 
\be
	C^{(\xi)}(r) =
\begin{cases}
& \left( 1-r^\xi\right) \mathbf 1_{r<1} \text{ for } \xi>0\\
&  \left(\log \frac{1}{r}\right)  \mathbf 1_{r<1}  \text{ for } \xi=0
\end{cases}.
\label{eq:piecewise}
\ee
The \corr subscript $\eta$ \rroc denotes a regularization over the  small-scale $\eta$,  in practice most easily defined using Fourier transforms \cite{supplemental}.  Please note that  the expressions \eqref{eq:piecewise} indeed  represent correlation functions for $0 < \xi \le 1$ \cite{yaglom1987correlation}, and we therefore restrict our analysis to this range. With this choice,  Eq. ~\eqref{eq:correlation} is then exactly and not just asymptotically satisfied. 
Spatial intermittency is modeled by the term $M_{\eta}^{(\gamma)} = e^{\gamma Y} W_1(dy,\cdot)$, namely the exponentiation of a regularized fractional Gaussian field $Y$ with vanishing Hurst exponent. This non-trivial operation  reguires to be suitably normalized  by the term $Z = e^{\gamma^2 \mathbb E (Y^2)} \sim \eta^{-\gamma^2}$. The mathematical expectation $\mathbb E$ denotes an average over \corr the random environment Y. \rroc  When $\gamma<\sqrt 2/2\simeq 0.707$, the limit  $ \eta \to 0$ then produces  a well-defined and non-trivial multifractal random distribution  called \emph{Gaussian multiplicative chaos}  \cite{chevillard2015peinture, pereira2016dissipative,rhodes2014gaussian,chevillard2019skewed} (later referred to as GMC). 
The multifractality prescribes the power-law scaling  $S_p(\ell) := \av{|u(x+r) - u(x)|^p}  \sim  C_p |r|^{\zeta(p)}$ in the inertial range  $\eta \le r \ll 1$, with quadratic variation  of the structure function exponents as
\be
	\begin{split}
	& \zeta(p) = p\left(\xi/2+\gamma^2\right) -\gamma^2p^2/2;
	\label{eq:expo}
	\end{split}
\ee
This is  a signature of log-normal multifractality  -- see Fig.~\ref{fig:1} for a numerical illustration using  Monte-Carlo averaging with $\eta=2^{-20}$, and  the power law extending over almost 5 decades. 
Here, the originality of the field \eqref{eq:MarkovGMC} comes from its temporal dependence. The Gaussian component is Markovian, and the random environment   	\eqref{eq:MarkovGMC} is analogous to a Kraichnan flow when we set the intermittency parameter $\gamma=0$. The GMC component is random but frozen-in-time: This feature will allow the spatial intermittency to play out in \eqref{eq:dynamics} even at the level of two-particle dynamics.\\

\paragraph{Two wrong \corr intuitive assumptions \rroc on multi-scaling \& spontaneous stochasticity.}
Eq.~\eqref{eq:expo} prescribes $\zeta(2)=\xi$, meaning that for the multifractal fields \eqref{eq:MarkovGMC}, 
the two-point correlation \eqref{eq:correlation}  is prescribed by the Hurst exponent, independently from the intermittency parameter $\gamma$.  Still, our multifractal fields  are rougher than their mono-fractal counterpart.
This is seen qualitatively from the numerical realizations of Fig.~\ref{fig:1} (c): Increasing $\gamma$ also increases the spikiness of the signal. More quantitatively,  the spatial roughness $\xi_K$  of a single realization of the random field $u^{\xi,\gamma}$ \corr is tied \rroc to its scaling exponents  $\zeta(p)$ through the  classical Kolmogorov continuity theorem  \cite{le2016brownian,evans2012introduction} as
\be
	\xi_K  =2  \sup_p \dfrac{\zeta(p)-1}{p}  \le \xi
	\label{eq:effectiveroughness}
\ee
For $\gamma=0$,  the mono-fractal behavior $\zeta(p) = \ p\xi/2$  holds for arbitrarily large $p$'s, and as such the \corr exponent \rroc $\xi_K$  identifies to the Hurst exponent $\xi$. For $\gamma \neq 0$, though, 
this value is reached at $p_K = \sqrt 2/\gamma$, which prescribes $\xi_K<\xi$ -- see Fig \ref{fig:1} (d).
On the one hand,  because of this enhanced roughness,  one could in principle expect that tracers advected in the intermittent fields \eqref{eq:dynamics} \corr are \rroc more likely to exhibit non-deterministic behavior than in Kraichnan flows, following the mantra ``The rougher, the more spontaneous  stochastic''.
On the other hand,  two-particle separations in the monofractal Kraichnan flows depend only on $\zeta(2) =\xi$ \cite{supplemental}: One may as well expect that at least at the level of two-particle separation, one should see no effect of intermittency.  We now present some formal  and  numerical results, 
to argue that none of the \corr intuitive assumptions formulated above  \rroc  are in fact correct.\\

\paragraph{From random fields to random potentials.}
We  focus on the dynamics of pair separations, obtained by considering $N=2$ in Eq.~\eqref{eq:dynamics}.
Similar to the Gaussian case \cite{frisch1998intermittency,gat1998anomalous},  tracers advected by Eq.~\eqref{eq:dynamics} can be interpreted in terms of particles interacting through a random  pairwise potential, and whose dynamics are prescribed  through the stochastic differential equation (SDE)
\be
	\label{eq:randompotentials}
\begin{split}
	& d X_i=
	\dfrac{1}{Z_\eta} \sum_{j=1,2}  L _{ij}\left(|X_1-X_2|\right) e^{\gamma Y(X_j)}W_j(dt)\\
	& \hspace{4cm}+\sqrt{2\kappa } B_i(dt)\\
& \text{with } L(r):=
\begin{pmatrix}
& 1 & 0\\
& 1-r^{\xi} &  \sqrt{1-(1-r^{\xi})^{2}}\\
\end{pmatrix}
\end{split}
 \ee

The matrix $L$ is a discrete analogue to the kernels $L_\eta^{(\xi)}$ featured in Eq.~\eqref{eq:MarkovGMC}, \corr except for \rroc the regularizing scale $\eta$. It corresponds to an explicit Choleski decomposition of the correlation matrix $C_2:=C(|X_i-X_j|)_{i,j=1,2}$ such that $LL^T=C_2$.
As a SDE version of the original dynamics  \eqref{eq:dynamics},  Eq.~\eqref{eq:randompotentials} comes with two advantages.
\emph{(i)} At a numerical level, it allows for Monte-Carlo sampling of trajectories without the need to generate the fields of Eq.~\eqref{eq:MarkovGMC} at each time-step, similar to the Gaussian setting \cite{frisch1998intermittency,gat1998anomalous}. 
 \emph{(ii)} At a formal level,  separation-time statistics can be obtained by means of  stochastic calculus and potential theory for Markov processes, in other words Feynman-Kac-like formulas. The word \emph{formal}  is advisory, as  the frozen-in-time GMC entering the dynamics could require cautious mathematical handling\cite{rhodes2014gaussian,garban2016liouville,rhodes2016lecture}, but this   goes way  beyond the  scope of the present letter.\\

\paragraph{The paradoxical interplay between intermittency and spontaneous stochasticity.}
Stochastic calculus suggests that for a prescribed realization of the GMC,  the  pair-separation time $T_1^Y(r)$ from scale $r$ to scale $1$  formally solves the boundary-value problem 
\be
\tag{$\gamma$-BVP}
\label{eq:pb2}
\begin{split}
	&  \mathcal L_{2}^Y T_1^Y = -1 \;\;\;\\
	&  \text{   with   } T_1^Y(1)=0 \text{ and } \partial_r T^{Y}_1(\eta)=0,
\end{split}
\ee
involving the GMC-dependent operator
\be
\label{eq:op2}
\begin{split}
	& \mathcal L_{2}^Y(X_1,r): =\dfrac{e^{2\gamma Y(X_1)}}{2Z_\eta^2} \left(r^\xi+ e^{2\gamma \Delta Y} \left(2-r^{\xi}\right)  \right)r^{\xi} \partial_{rr},
\end{split}
\ee
which features 
the increment $\Delta Y:= Y\left(X_1+r\right)-Y\left(X_1\right)$.
For $\gamma \neq 0$, \eqref{eq:pb2} features a non-trivial coupling between the pair-separation time and the underlying GMC. Because of this coupling, \eqref{eq:pb2} is not closed and one cannot a priori solve it explicitly for $T_1^Y$.
\corr Setting  $\gamma=0$ retrieves the Gaussian setting and provides a statistical decoupling\cite{gawedzki2008soluble,supplemental}, which makes Eq.~\eqref{eq:pb2}-\eqref{eq:op2} solvable. \rroc
For $\gamma \neq 0$, we define the annealed separation-time
\be
	\tau_1:= \avY{T_1^Y},
\ee
where we recall that the expectation $\avY{\cdot}$ denotes an average over the  GMC random environment.
A non-trivial decoupling is then obtained under the mean-field Ansatz
\be
	\label{eq:meanfield}
	\avY{e^{\gamma \Delta Y} T_1^Y}= \avY{e^{\gamma \Delta Y}} \tau_1.
\ee
For  $\eta \ll 1$, \eqref{eq:pb2} under Anzatz \eqref{eq:meanfield} formally becomes
\be
\begin{split}
	&  \mathcal L^*_2 \tau_1 = -1 \;\;\; \text{   with   } \tau_1(1)=0 \text{ and } \partial_r \tau_1(0)=0,\\
	& \text{ for } 	\mathcal L_{2}^*(r): = \left(1-\dfrac{r^\xi}{2}\right) r^{\xi+4\gamma^2} \partial_{rr}.
\end{split}
\label{eq:op3}
\ee
We refer the reader to Supplemental Material \cite{supplemental} for details on  derivations of  
\eqref{eq:op2}-\eqref{eq:op3} . As for now, we observe that   the term $(1-r^\xi/2)$ is bounded and of order $O(1) $. Hence, the separation times behave as if the multifractal random flows were   Gaussian, yet with effective \emph{driving} Hurst parameter 
\be
	\label{eq:xigamma}
	\xi_{\gamma} := \xi + 4 \gamma^2 > \xi>\xi_K.  
\ee
This Gaussian flow is smoother than the flow at $ \gamma =0$~! This calculation evidences a highly paradoxical interplay between multifractality, roughness and spontaneous stochasticity:  Increasing $\gamma$ makes the flow rougher in terms of the  effective roughness $\xi_K$ deduced from Kolmogorov theorem, but makes the flow smoother in terms of the spontaneous stochasticity of tracers,
driven by $\xi_\gamma$ given above. A practical consequence of Eq.~\eqref{eq:xigamma} is the presence of a phase transition driven by $\gamma$,  and characterized by $\xi_\gamma=1$. This  prescribes the mean-field critical curve
\be
	\label{eq:xigamma_critical}
	\gamma_c= \dfrac{1}{2}\sqrt{1-\xi}:
\ee
For $0<\xi<1$, tracers are spontaneously stochastic when  $\gamma<\gamma_c $ and deterministic when $\gamma\ge \gamma_c$.
For $\xi =1$,  the Gaussian case is deterministic, and the critical $\gamma_c$ is vanishing, as should be. For $\xi =0$, this prescribes $\gamma_c=1/2$,  less than  the maximum value $\sqrt{2}/2$ allowed for the GMC. The critical value $\gamma_c$ can therefore in principle be achieved for any Hurst exponent $ \xi \in ]0,1]$,\\

\paragraph{Numerics.}
To illustrate the  rationale of the  prediction \eqref{eq:xigamma} and  mean-field Ansatz \eqref{eq:meanfield}, we now report results of  Monte-Carlo sampling of pair trajectories, obtained from   two different methods, and where we vary the  levels of roughness $\xi$, intermittency $\gamma$ and the regularization scale $\eta$.
The first method is field-based. It uses direct integration of the dynamics \eqref{eq:dynamics}  with standard Euler-Maruyama method, and requires to generate a new spatial realization of the field \eqref{eq:MarkovGMC} at each timestep.  Tracers are then advected by smoothly interpolating the velocity field at their current positions. The second method is SDE-based. It uses the representation \eqref{eq:randompotentials} in terms of interacting particles and only requires to generate a single   field, namely the frozen  GMC, per a pair of trajectories.  In the SDE setting, in order to enhance numerical stability,  we add  a callback function  ensuring exact reflecting boundary conditions  for particles reaching $\eta$. \\
Beyond the physical parameters $\xi,\gamma,\eta$,  both methods require to set values for the field resolutions $N$, \corr the number of  trajectory realizations\rroc, the   timesteps $dt$ -- Table \ref{tab:param} lists  the  essential numerical parameters.  To ensure a finite-time completion of the numerical algorithms,  we use a maximal time $T_{max}$ over which the numerics are stopped: Our Monte-Carlo sampling  therefore does  not measure $\tau_1(\eta)$ but rather the estimate $\tilde \tau_1= \av{T_1^Y \wedge   T_{max}}$. 
\begin{table}
\begin{tabular}{
|c|c|c|c|c|c|c|}
 \hline
 $\xi$& $\gamma$& $N$ & $\eta$  & $dt$ & $T_{max}$&Realizations\\
\hline
 $ 2/9 $ to  $1$ &$ 0$ to  $0.6$& $ 2^{7}$ to  $2^{14}$ & $4\pi/N$ & $10^{-4}$ & $8$ to $64$ &$10^5$\\
\hline
\end{tabular}
\caption{Numerical parameters for Fig.\ref{fig:2} and Fig. \ref{fig:3} for both methods. For SDE-based numerics, realizations are independent while \corr for the \rroc field-based numerics, we use minibatches of 100 trajectories to reach $10^{5}$ samples. }
\label{tab:param}
\end{table}

\begin{figure}
\includegraphics[width=0.49\columnwidth,trim=2cm 0cm 0cm 0cm, clip]{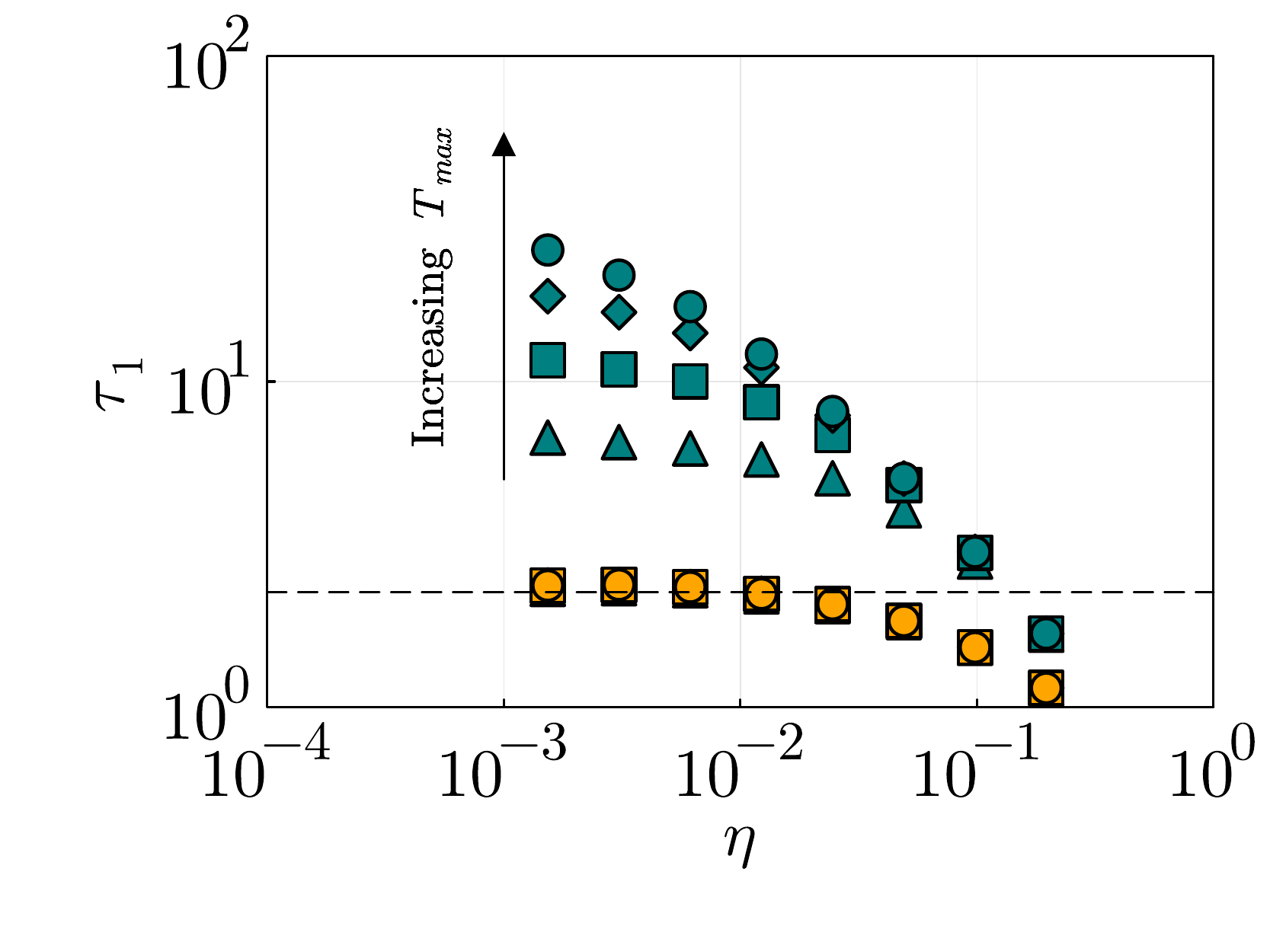}
\includegraphics[width=0.49\columnwidth,trim=2cm 0cm 0.cm 0cm, clip]{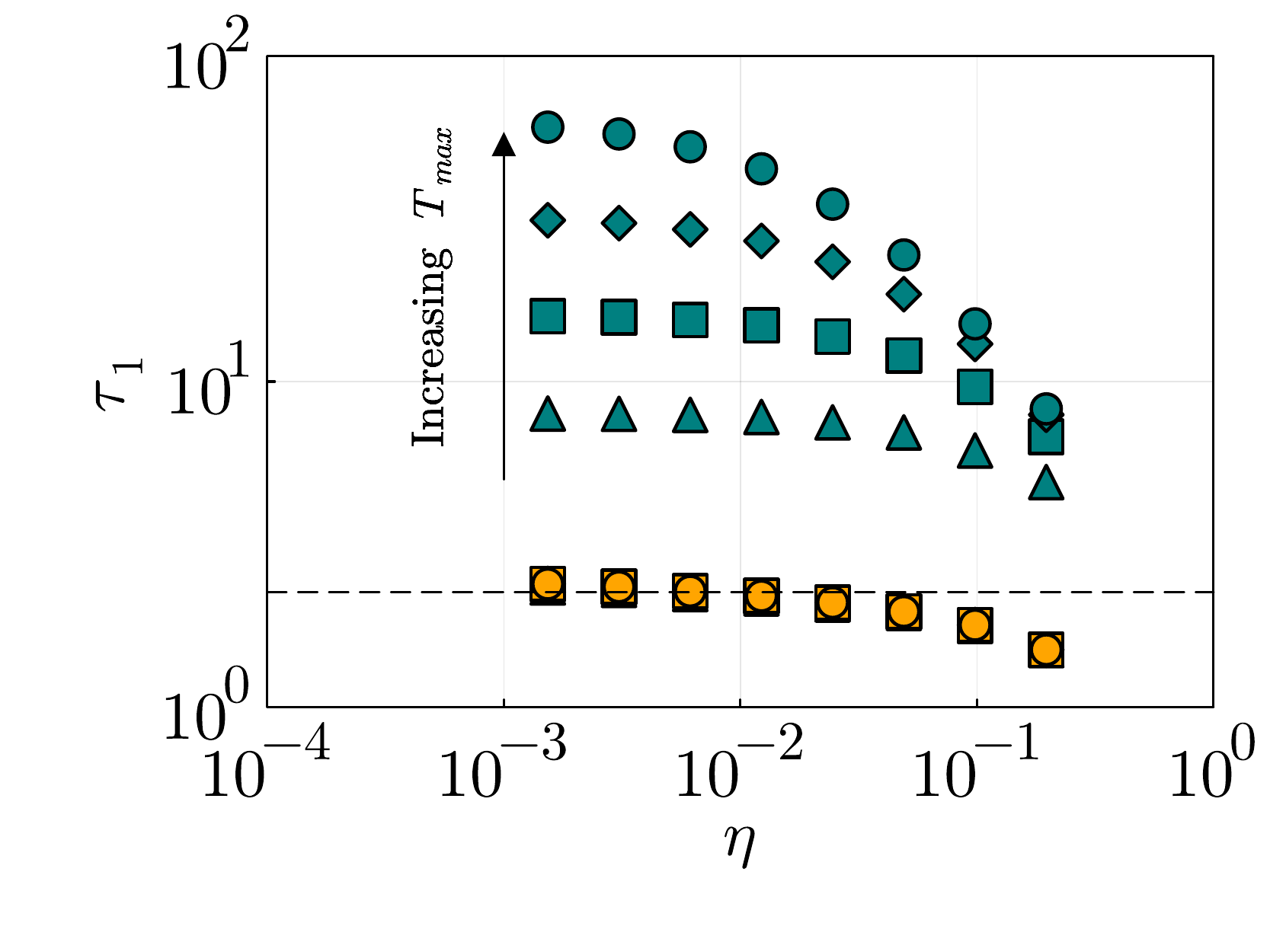}\\
\includegraphics[width=0.49\columnwidth,trim=1cm 0cm 0cm 0cm, clip]{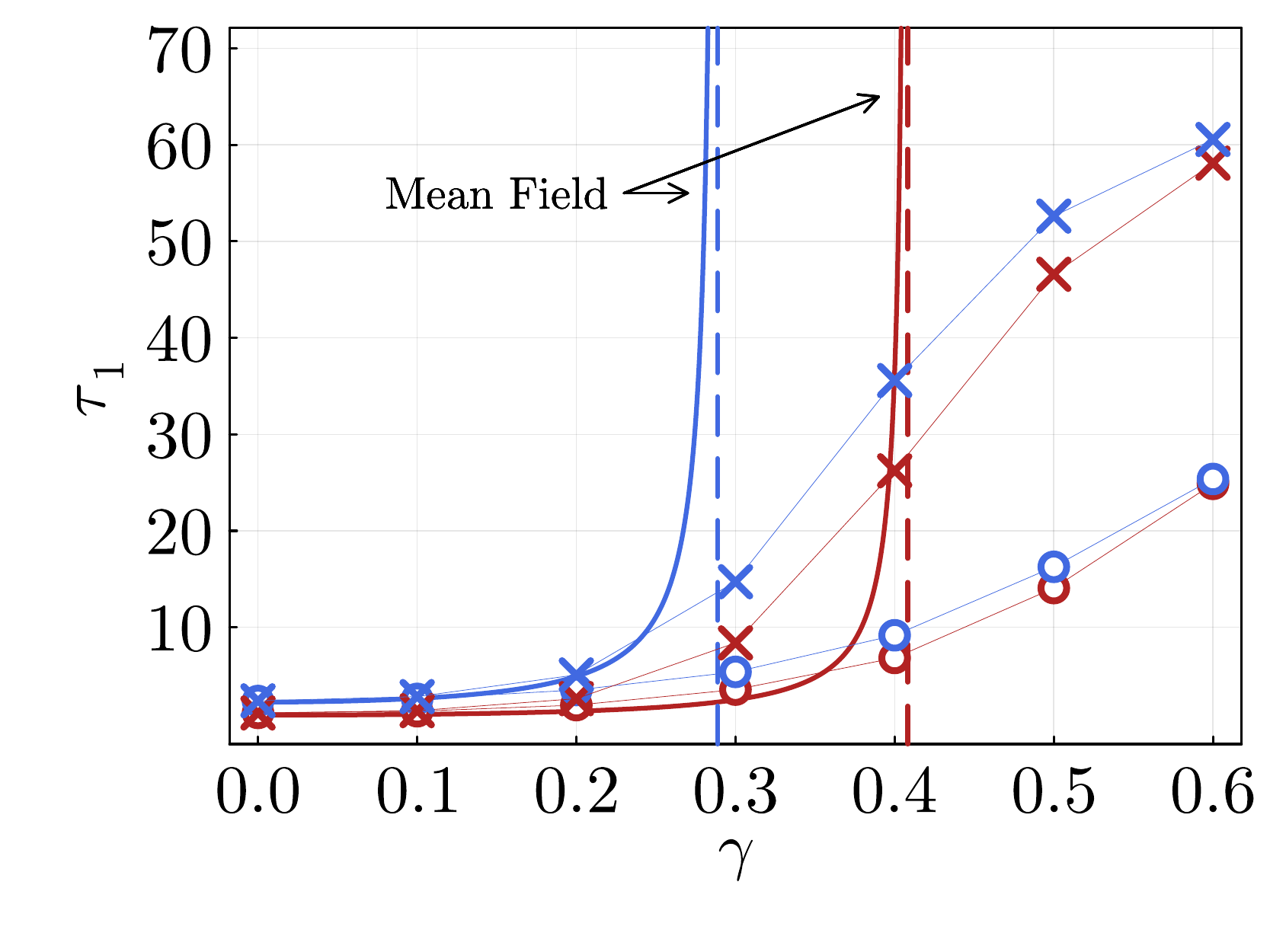}
\includegraphics[width=0.49\columnwidth,trim=0cm 0cm 0.3cm 0cm, clip]{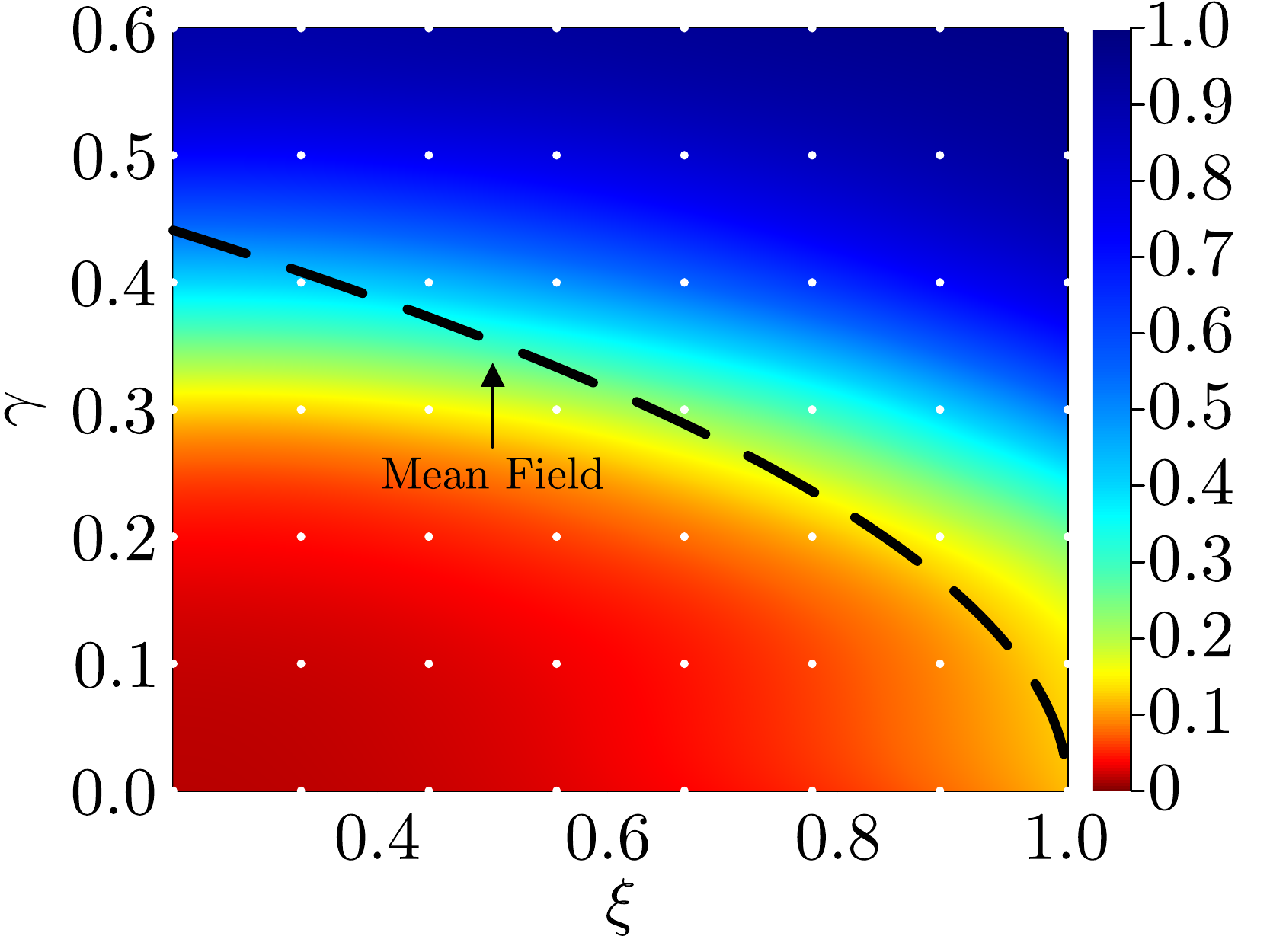}
\setlength{\unitlength}{\columnwidth}
\begin{picture}(1,0)(0,0)
\put(0.12,0.54){(a)}
\put(0.62,0.54){(b)}
\put(0.1,0.16){(c)}
\put(0.6,0.16){(d)}
\end{picture}

\caption{  
(a) Convergence with $\eta$ of the separation time  at $\xi=2/3$  for field-based numerics, using maximal simulation times  $T_{max}=8 (\triangle),16(\square),32 (\diamond),64 (\circ)$,  for $\gamma =0$ (orange) and $\gamma =0.6$ (green). Dashed line indicates the Kraichnan flow value. (b) Same but for SDE-based  numerics.
(c)  $\tau_1$ against $\gamma$ at $\xi=1/3$ (red) and $2/3$ (blue) for  field-based ($\circ$) and SDE-based ($\times$) numerics, using the smallest $\eta=10^{-3}$ and largest $T_{max}=64$. (d)
Colormap for the average separation time $\tau_1$   against the roughness and intermittency parameters $\xi,\gamma$ from SDE-based numerics. Data are normalized by  $T_{max}=64$, and interpolated from that measured at the white dots. Red (blue) rendering indicate values  close to $0$ (1) suggestive of spontaneously stochastic (deterministic) behavior. }
\label{fig:2}
\end{figure}
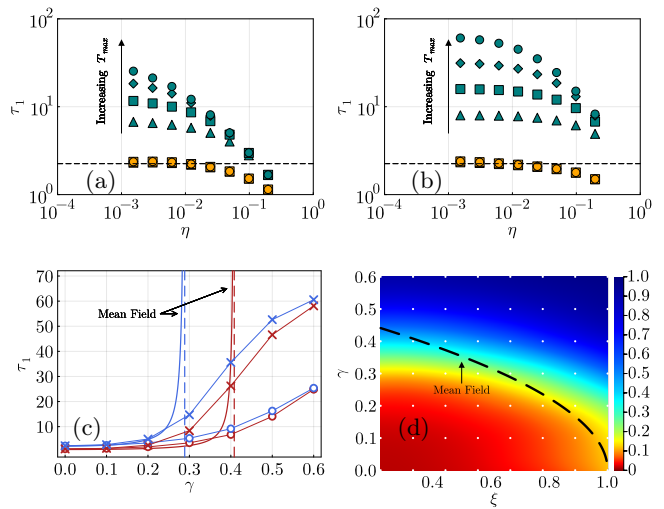

\begin{figure}
\includegraphics[width=0.49\columnwidth,trim=1cm 0cm 1cm 0cm, clip]{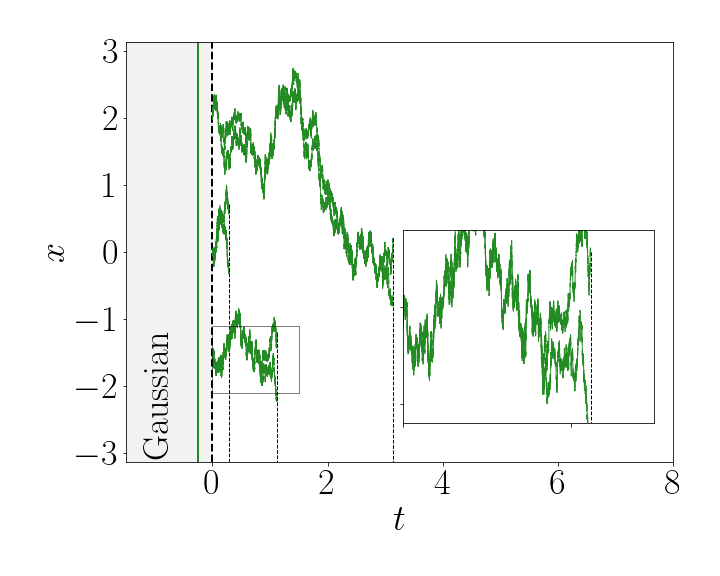}
\includegraphics[width=0.49\columnwidth,trim=1cm 0cm 1cm 0cm, clip]{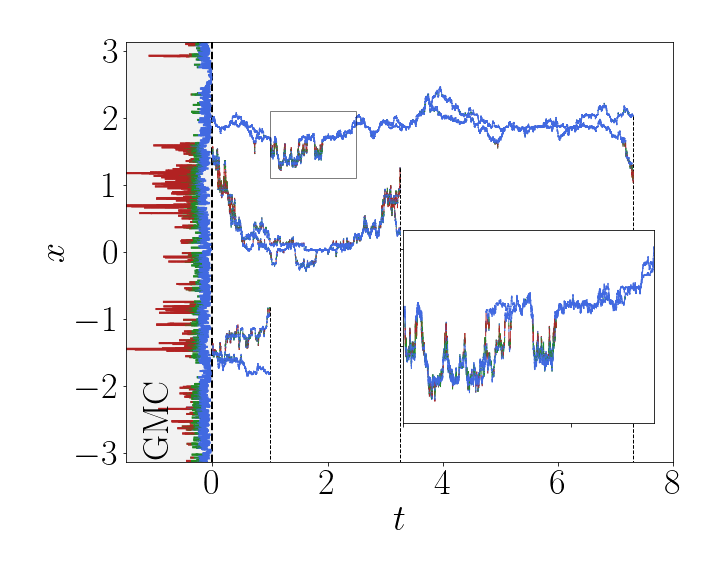}\\
\includegraphics[width=0.49\columnwidth,trim=1.4cm 0cm 0cm 0cm, clip]{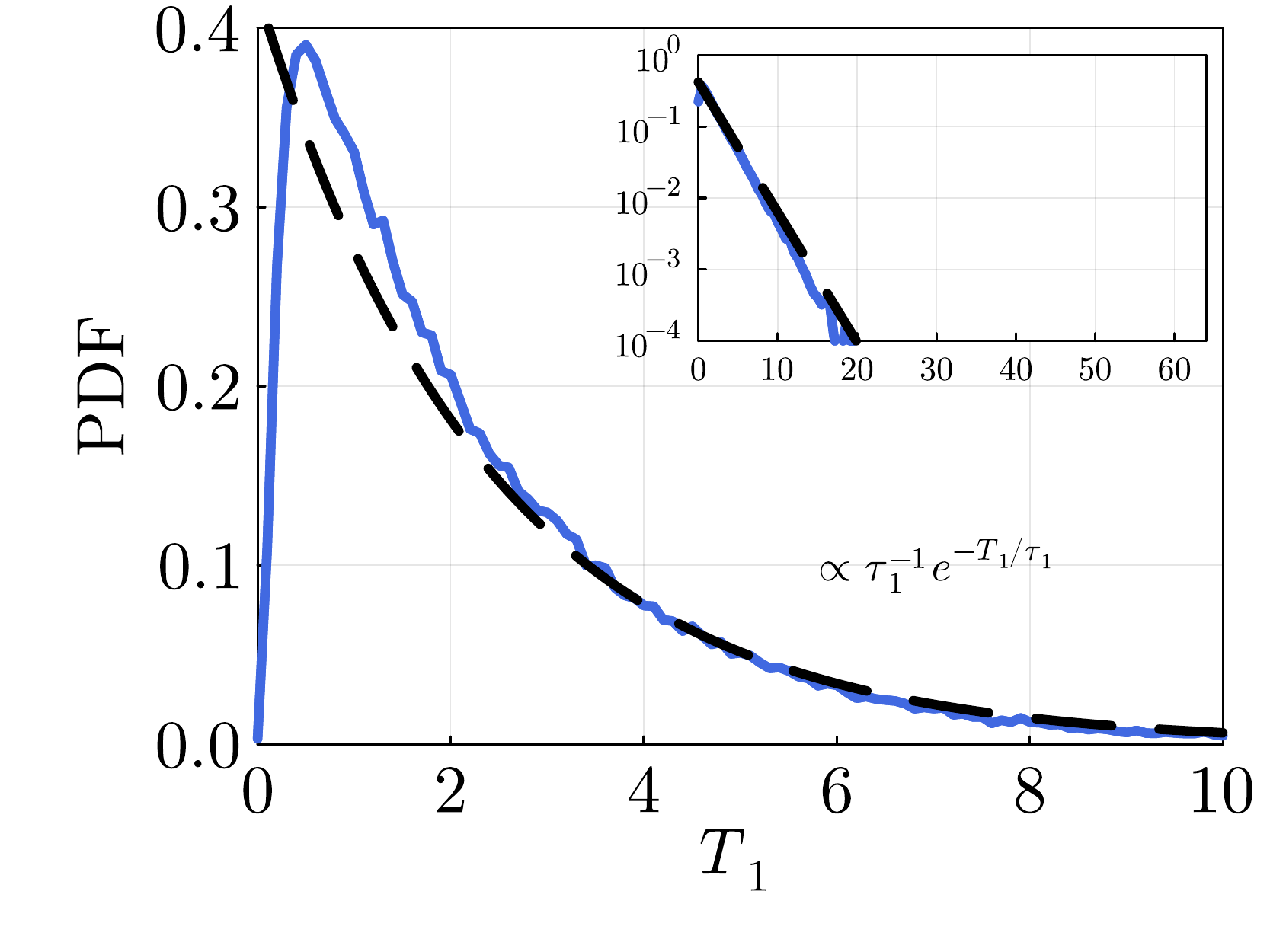}
\includegraphics[width=0.49\columnwidth,trim=1.4cm 0cm 0cm 0cm, clip]{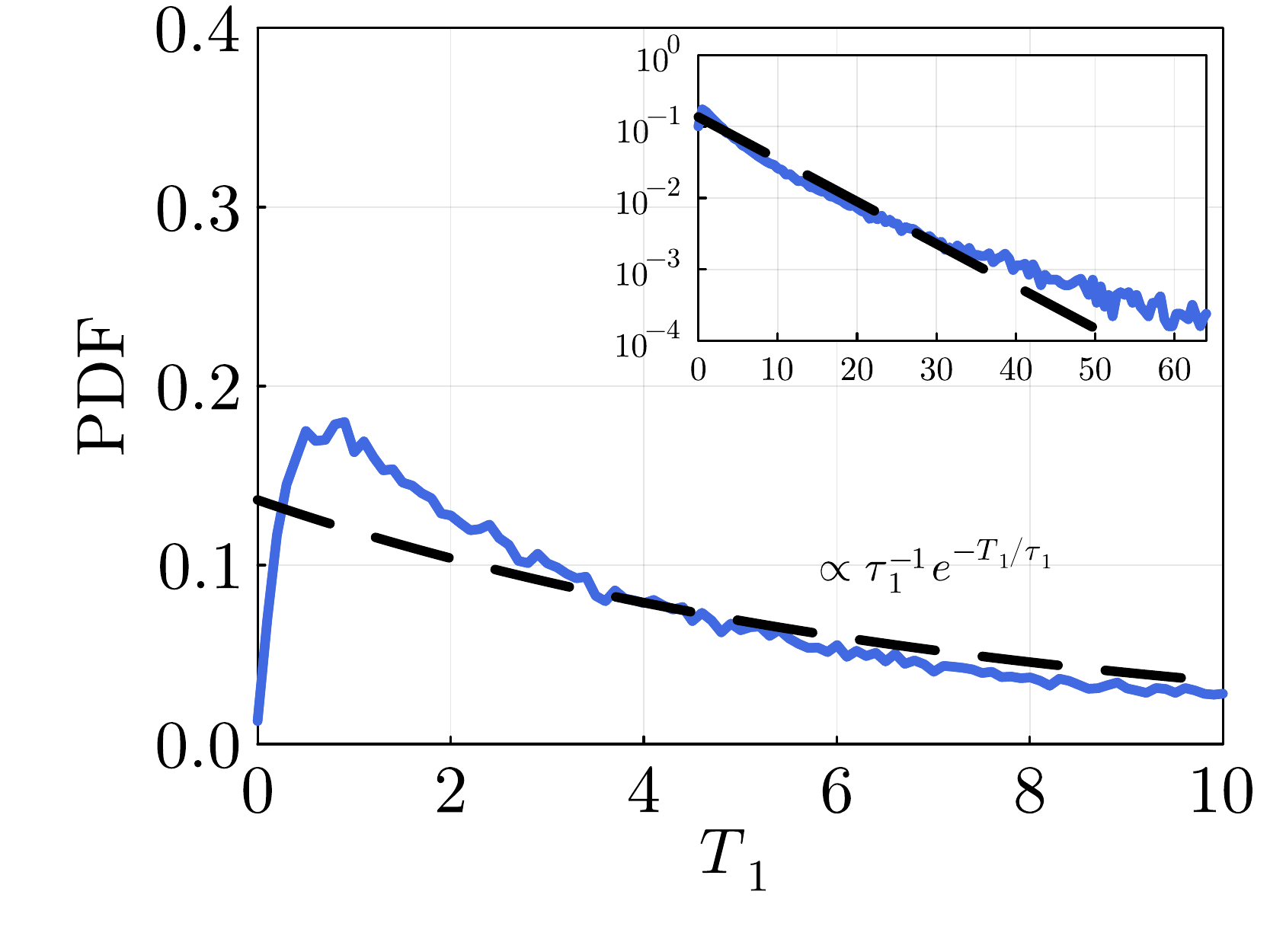}
\setlength{\unitlength}{\columnwidth}
\begin{picture}(1,0)(0,0)
\put(0.43,0.55){(a)}
\put(0.92,0.55){(b)}
\put(0.43,0.16){(c)}
\put(0.92,0.16){(d)}
\end{picture}
\caption{
 (a) 3 Random realizations of initially coalesced pair trajectories until their separation times $T_1$ for    driving Hurst parameter $\xi_\gamma=2/3$ with  $\xi=2/3, \gamma= 0$. 
 (b) Same but with $\xi=1/3, \gamma= \sqrt 3/6.$  Color indicates the magnitude of the underlying GMC, whose profile is represented vertically in the negative axis.
(c)  PDF of separation-times obtained from SDE-based numerics  for  $\xi=2/3, \gamma= 0$.  Inset uses log-scale for the $y$-axis.  (b) Same but for $\xi=1/3, \gamma= \sqrt 3/6$.}
\label{fig:3}
\end{figure}
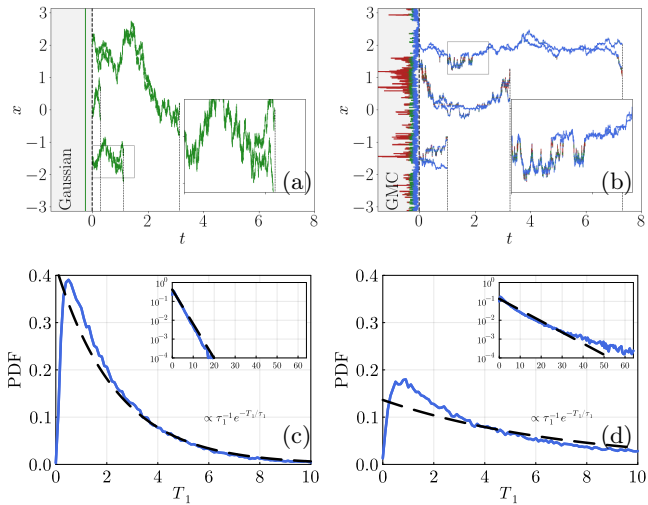

Fig.~\ref{fig:2} and  Fig.~\ref{fig:3} summarize our numerical observations, and those  prove compatible with the mean-field predictions.
As seen from Fig.~\ref{fig:2},   we monitor two types of behaviors when prescribing  $0<\xi<1$.
Setting for instance $\xi=2/3$  as in Panels (a,b), we observe that for small  $\gamma $,  the estimates  $\tilde \tau_1$ converge to finite value, independent from $T_{max}$. Upon decreasing $\eta$, it is found that this value is compatible with the mean-field prediction $\tau_1^{mf}(\xi_\gamma) =\left(2(1-\xi_\gamma)(1-\xi_\gamma/2)\right)^{-1} $ involving the Kraichnan flow estimate with the  driving Hurst exponent as input \cite{supplemental}. This   evidences the spontaneous stochastic nature of separations   for small $\gamma$.
For large $\gamma$, the apparent convergence of $\tilde \tau_1$ when decreasing $\eta$ is a numerical artifact, as the limiting value grows with  $T_{max}$. This signals deterministic behavior,  with the particles not separating in the limit $\eta \to 0$.
As such, for all the \corr values of $\xi$ considered in this work\rroc , our numerics reflect the presence of a  phase transition at finite value of $\gamma$.
This is in agreement with our  mean-field argument, and  substantiates our claim that intermittency here favors deterministic behavior.  
The onset of deterministic behavior when increasing $\gamma$ is found to be similar when using either  field-based or SDE-based numerics. Focusing on the cases  $\xi=1/3$ and $\xi=2/3$, we find good compatibility with the   mean-field prediction in the latter case and observe deviations in the former case -- see Panel (c). As seen from Panel (d),  the mean-field prediction accurately captures the transition between deterministic and non-deterministic behaviors for small  values of the effective roughness, corresponding to the larger values of $\xi$. The  
agreement seems to deteriorate for  smaller $\xi$. 
This discrepancy suggests that \corr the mean-field approach  becomes inaccurate  in the latter regime\rroc,  but one cannot rule out  a defect of the numerics : As $\xi \to 0^+$, the paths become very rough, and  the Euler-Maruyama scheme may become unfit even when combined with  very \corr fine \rroc timestepping.

Fig.~\ref{fig:3} shows the effect of non-Gaussianity playing out at fixed value of the driving Hurst exponent $\xi_\gamma=2/3$ in the  spontaneous stochastic regime. As seen from Panels (a) and (b), the  behaviors between  Gaussian and non-Gaussian settings are qualitatively different, although by construction, both share  the same mean-field average separation time. In the Gaussian case, the GMC $\propto e^{\gamma Y}$ is unity, and  the behavior of pairs is statistically independent  from their absolute positions. In the non-Gaussian case, the GMC is non-trivial  and  we observe \corr a dependence on \rroc the local values of the
magnitude $\gamma Y$.  This  is compatible with the fact that  Eq.~\eqref{eq:op2} ruling the pair separations is not closed, unless one further averages over the GMC realizations. This behavior reflects in the PDF of exit-times. At large times, the Gaussian setting exhibits the exponential decay  $\propto e^{-t/\tau_1}$  predicted by the Kraichnan flow theory \cite{supplemental}. The non-Gaussian case deviates from the exponential behavior; it exhibits fat tails,  likely reflecting particles  trapped in quiet ``valleys'' of the frozen-in-time GMC. Understanding the details of this slow decay requires tools more refined than the present mean-field approach and is left for future studies. \\

\paragraph{Concluding remarks.}
We have proposed a non-trivial extension of the Kraichnan flow theory towards a multifractal setting, that we obtained by decorating the original  Markovian Gaussian  flows  with  a  frozen-in-time Gaussian multiplicative chaos. 
Multifractality makes the flow rougher in terms of the Kolmogorov roughness $\xi_K$, but the spontaneous stochasticity of two-particle separation  maps to that of a smoother Gaussian environment, with Hurst exponent $\xi_\gamma> \xi>\xi_K$. This  paradoxical effect is all the less intuitive,  as the second-order structure functions of our parametric family of fields are  characterized by constant $\zeta(2)=\xi$ independent  of the level of the intermittency. \corr This is  an example  of  a ``smoother ride over a rougher sea'' at play in scalar transport\cite{frisch2014very,falkovich2001particles}, and possibly connected to the mathematical theory of regularization by noise \cite{galeati2021noiseless}. \rroc 
The availability of a SDE-based interpretation  suggests a new playground to build a fundamental understanding of transport in multifractal environments. \corr This includes tackling higher-dimensions, revisiting scalar intermittency and connections with anomalous dissipation \cite{drivas2017lagrangian,valade2023anomalous}, or more generally addressing irreversibility\cite{bauer1999sailing,eyink2015spontaneous} and  universality of transport.  On the latter aspect, let us here point out \rroc that  the specific value of the driving roughness $\xi_\gamma$ is clearly model-dependent. \corr For example,  using white-in-time versions of the GMC (not shown here), both the mean-field approach and  the numerics   yield $ \xi_\gamma=\xi$\rroc. The dependence upon time-correlation is reminiscent of that observed in the Gaussian case  \cite{chaves2003lagrangian}. However, one could  expect that the feature $\xi_\gamma \ge \zeta(2)$ is a robust feature of pair-dispersion in multifractal environments, and this could prove
crucial  in assessing transport in genuine turbulent environments. Roughness and intermittency levels measured in 3D homogeneous isotropic turbulence corresponds to  $\xi\simeq 0.69$ and $\gamma\simeq 0.15$\cite{dubrulle1994intermittency, chevillard2019skewed,mailybaev2022hidden}. In our model, this choice of parameters results in driving Hurst parameter $\xi_\gamma \simeq 0.77$. Transposing this exponent to a deterministic setting would yield  Richardson diffusion  $R \sim t^{1.62}$, \corr noticeably \rroc different from  $t^{3/2}$. 
Naturally, while they extend the Kraichnan flows to a  more realistic setting, our  unidimensional multifractal environments remain  caricatures of  genuine turbulent fields, and lack  essential ingredients such  as skewness, temporal correlations, incompressiblity, etc. At a more mathematical level,  our analysis  suggests that \corr the problem of transport \rroc in multifractal random flows, be them Eq.~\eqref{eq:MarkovGMC} or variations thereof, might prove solvable. Here,  the use of a  frozen-in-time GMC as a random environment is strongly reminiscent of the parabolic Anderson model used in condensed matter physics \cite{konig2015parabolic} and the  Liouville Brownian motion entering the construction of field theories in  the context of 2D  quantum gravity \cite{rhodes2016lecture,garban2016liouville}.  Exploiting those analogies may  provide  a path towards a rigorous treatment of transport  in multifractal turbulent-like random environments,
and quantitative modeling of turbulent transport in terms of random fields.

\section*{Acknowledgments}
We thank A. Barlet, A. Cheminet, B. Dubrulle \& A. Mailybaev   for continuing discussions. ST acknowledges support 
 from the French-Brazilian network in Mathematics for visits  at Impa in Southern summers 2022 \& 2023, where this work was initiated, 
and  thanks L. Chevillard for  essential insights on multifractal random fields.

\bibliographystyle{unsrt}
\bibliography{Biblio}

\begin{thebibliography}{10}

\bibitem{richardson1926atmospheric}
L.F. Richardson.
\newblock Atmospheric diffusion shown on a distance-neighbour graph.
\newblock {\em Proc. Roy. Soc. Lond.}, 110(756):709--737, 1926.

\bibitem{jullien1999richardson}
M.-C. Jullien, J.~Paret, and P.~Tabeling.
\newblock Richardson pair dispersion in two-dimensional turbulence.
\newblock {\em Phys. Rev. Lett.}, 82(14):2872, 1999.

\bibitem{boffetta2002statistics}
G.~Boffetta and I.~Sokolov.
\newblock Statistics of two-particle dispersion in two-dimensional turbulence.
\newblock {\em Phys. Fluids}, 14(9):3224--3232, 2002.

\bibitem{boffetta2002relative}
G.~Boffetta and I.~Sokolov.
\newblock Relative dispersion in fully developed turbulence: the
  {R}ichardson’s law and intermittency corrections.
\newblock {\em Phys. Rev. Lett.}, 88(9):094501, 2002.

\bibitem{bitane2012time}
R.~Bitane, H.~Homann, and J.~Bec.
\newblock Time scales of turbulent relative dispersion.
\newblock {\em Phys. Rev. E}, 86(4):045302, 2012.

\bibitem{thalabard2014turbulent}
S.~Thalabard, G.~Krstulovic, and J.~Bec.
\newblock Turbulent pair dispersion as a continuous-time random walk.
\newblock {\em J. Fluid Mech.}, 755:R4, 2014.

\bibitem{bourgoin2015turbulent}
M.~Bourgoin.
\newblock Turbulent pair dispersion as a ballistic cascade phenomenology.
\newblock {\em J. Fluid Mech.}, 772:678--704, 2015.

\bibitem{buaria2015characteristics}
D.~Buaria, B.~Sawford, and P-K. Yeung.
\newblock Characteristics of backward and forward two-particle relative
  dispersion in turbulence at different {R}eynolds numbers.
\newblock {\em Phys. Fluids}, 27(10):105101, 2015.

\bibitem{tan2022universality}
S.~Tan and R.~Ni.
\newblock Universality and intermittency of pair dispersion in turbulence.
\newblock {\em Phys. Rev. Lett.}, 128(11):114502, 2022.

\bibitem{buaria2023comment}
D.~Buaria.
\newblock Comment on ``{U}niversality and intermittency of pair dispersion in
  turbulence''.
\newblock {\em Phys. Rev. Lett.}, 130:029401, 2023.

\bibitem{lorenz69predictability}
E.~Lorenz.
\newblock The predictability of a flow which possesses many scales of motion.
\newblock {\em Tellus}, 21(3):289--307, 1969.

\bibitem{mailybaev2016spontaneously}
A.~Mailybaev.
\newblock Spontaneously stochastic solutions in one-dimensional inviscid
  systems.
\newblock {\em Nonlinearity}, 29(8):2238, 2016.

\bibitem{mailybaev2016spontaneous}
A.~Mailybaev.
\newblock Spontaneous stochasticity of velocity in turbulence models.
\newblock {\em Mult. Mod. \& Sim.}, 14(1):96--112, 2016.

\bibitem{thalabard2020butterfly}
S.~Thalabard, J.~Bec, and A.~Mailybaev.
\newblock From the butterfly effect to spontaneous stochasticity in singular
  shear flows.
\newblock {\em Comm. Phys.}, 3(1):1--8, 2020.

\bibitem{mailybaev2022spontaneously}
A.~Mailybaev and A.~Raibekas.
\newblock Spontaneously stochastic {A}rnold’s cat.
\newblock {\em Arnold Math. Jour.}, 2022.

\bibitem{mailybaev2022spontaneous}
A.~Mailybaev and A.~Raibekas.
\newblock Spontaneous stochasticity and renormalization group in discrete
  multi-scale dynamics.
\newblock {\em arXiv preprint arXiv:2207.06158}, 2022.

\bibitem{gawedzki2008soluble}
K.~Gaw\k{e}dzki.
\newblock Soluble models of turbulent transport.
\newblock In {\em Non-equilibrium statistical mechanics and turbulence}, number
  355. 2008.

\bibitem{e2000generalized}
W.~E and E.~Vanden~Eijnden.
\newblock Generalized flows, intrinsic stochasticity, and turbulent transport.
\newblock {\em Proc. Nat. Acad. Sci. U.S.A.}, 97:8200--8205, 2000.

\bibitem{kupiainen2003nondeterministic}
A~Kupiainen.
\newblock Nondeterministic dynamics and turbulent transport.
\newblock {\em Ann. H. Poincar\'{e}}, 4(2):713--726, 2003.

\bibitem{chaves2003lagrangian}
M.~Chaves, K.~Gaw\k{e}dzki, P.~Horvai, A.~Kupiainen, and M.~Vergassola.
\newblock Lagrangian dispersion in {G}aussian self-similar velocity ensembles.
\newblock {\em J. Stat. Phys.}, 113(5):643--692, 2003.

\bibitem{falkovich2001particles}
G.~Falkovich, K.~Gawedzki, and M.~Vergassola.
\newblock Particles and fields in fluid turbulence.
\newblock {\em Rev. Mod. Phys.}, 73(4):913, 2001.

\bibitem{le2002integration}
Y.~Le~Jan and O.~Raimond.
\newblock Integration of {B}rownian vector fields.
\newblock {\em Ann. Probab.}, 30(2):826--873, 2002.

\bibitem{le2004flows}
Y.~Le~Jan and O.~Raimond.
\newblock Flows, coalescence and noise.
\newblock {\em Ann. Probab.}, 32(2):1247--1315, 2004.

\bibitem{gawedzki2001turbulent}
K.~Gaw\k{e}dzki.
\newblock Turbulent advection and breakdown of the {L}agrangian flow.
\newblock In {\em Intermittency in Turbulent Flows}.

\bibitem{gawedzki2000phase}
K.~Gaw\k{e}dzki and M.~Vergassola.
\newblock Phase transition in the passive scalar advection.
\newblock {\em Phys. D: Nonlin. Phen.}, 138:63--90, 2000.

\bibitem{supplemental}
See {S}upplemental {M}aterial for a short reminder on separation times in
  {K}raichnan flow theory, together with explicit details on both the
  field-based numerics and mean-field calculations.

\bibitem{reneuve2020flow}
J.~Reneuve and L.~Chevillard.
\newblock Flow of spatiotemporal turbulent-like random fields.
\newblock {\em Phys. Rev. Lett.}, 125(1):014502, 2020.

\bibitem{yaglom1987correlation}
A.~Yaglom.
\newblock {\em Correlation Theory of Stationary and Related Random Functions,
  Volume I: Basic Results}, volume 131.
\newblock Springer, 1987.

\bibitem{chevillard2015peinture}
L.~Chevillard.
\newblock {\em Une peinture al{\'e}atoire de la turbulence des fluides}.
\newblock {HDR} thesis, ENS Lyon, 2015.

\bibitem{pereira2016dissipative}
R.~Pereira, C.~Garban, and L.~Chevillard.
\newblock A dissipative random velocity field for fully developed fluid
  turbulence.
\newblock {\em J. Fluid Mech.}, 794:369--408, 2016.

\bibitem{rhodes2014gaussian}
R.~Rhodes and V.~Vargas.
\newblock Gaussian multiplicative chaos and applications: a review.
\newblock {\em Probab. Surveys}, 11:315--392, 2014.

\bibitem{chevillard2019skewed}
L.~Chevillard, C.~Garban, R.~Rhodes, and V.~Vargas.
\newblock On a skewed and multifractal unidimensional random field, as a
  probabilistic representation of {K}olmogorov’s views on turbulence.
\newblock In {\em Ann. H. Poincaré.}, volume~20, pages 3693--3741. Springer,
  2019.

\bibitem{le2016brownian}
J.-F. Le~Gall.
\newblock {\em Brownian motion, martingales, and stochastic calculus}.
\newblock Springer, 2016.

\bibitem{evans2012introduction}
L.~Evans.
\newblock {\em An introduction to stochastic differential equations},
  volume~82.
\newblock Am. Math. Soc., 2012.

\bibitem{frisch1998intermittency}
U.~Frisch, A.~Mazzino, and M.~Vergassola.
\newblock Intermittency in passive scalar advection.
\newblock {\em Phys. Rev. Lett.}, 80(25):5532, 1998.

\bibitem{gat1998anomalous}
O.~Gat, I.~Procaccia, and R.~Zeitak.
\newblock Anomalous scaling in passive scalar advection: {M}onte-{C}arlo
  {L}agrangian trajectories.
\newblock {\em Phys. Rev. Lett.}, 80(25):5536, 1998.

\bibitem{garban2016liouville}
C.~Garban, R.~Rhodes, and V.~Vargas.
\newblock Liouville {B}rownian motion.
\newblock {\em Ann. Probab.}, pages 3076--3110, 2016.

\bibitem{rhodes2016lecture}
R.~Rhodes and V.~Vargas.
\newblock Lecture notes on {G}aussian multiplicative chaos and {L}iouville
  quantum gravity.
\newblock {\em arXiv preprint arXiv:1602.07323}, 2016.

\bibitem{frisch2014very}
U.~Frisch and V.~Zheligovsky.
\newblock A very smooth ride in a rough sea.
\newblock {\em Comm. Math. Phys.}, 326:499--505, 2014.

\bibitem{galeati2021noiseless}
L.~Galeati and M.~Gubinelli.
\newblock Noiseless regularisation by noise.
\newblock {\em Rev. Mat. Ib.}, 38(2):433--502, 2021.

\bibitem{drivas2017lagrangian}
T.~Drivas and G.~Eyink.
\newblock A {L}agrangian fluctuation--dissipation relation for scalar
  turbulence. part i. flows with no bounding walls.
\newblock {\em J. Fluid Mech.}, 829:153--189, 2017.

\bibitem{valade2023anomalous}
N.~Valade, S.~Thalabard, and J.~Bec.
\newblock Anomalous dissipation and spontaneous stochasticity in deterministic
  surface quasi-geostrophic flow.
\newblock In {\em Ann. H. Poincar{\'e}}, pages 1--23. Springer, 2023.

\bibitem{bauer1999sailing}
M.~Bauer and D.~Bernard.
\newblock Sailing the deep blue sea of decaying {B}urgers turbulence.
\newblock {\em J. Phys. A: Math. Gen.}, 32(28):5179, 1999.

\bibitem{eyink2015spontaneous}
G.~Eyink and T.~Drivas.
\newblock Spontaneous stochasticity and anomalous dissipation for {B}urgers
  equation.
\newblock {\em J. Stat. Phys.}, 158:386--432, 2015.

\bibitem{dubrulle1994intermittency}
B.~Dubrulle.
\newblock Intermittency in fully developed turbulence: {L}og-{P}oisson
  statistics and generalized scale covariance.
\newblock {\em Phys. Rev. Lett.}, 73(7):959, 1994.

\bibitem{mailybaev2022hidden}
A.~Mailybaev and S.~Thalabard.
\newblock Hidden scale invariance in {N}avier--{S}tokes intermittency.
\newblock {\em Phil. Trans. Roy. Soc. A}, 380(2218):20210098, 2022.

\bibitem{konig2015parabolic}
W.~K{\"o}nig and T.~Wolff.
\newblock The parabolic {A}nderson model.
\newblock {\em Preprint. Available at www. wiasberlin. de/people/koenig}, 2015.

\end{thebibliography}


\begin{thebibliography}{10}

\bibitem{gawedzki2002soluble}
K.~Gaw\k{e}dzki.
\newblock Soluble models of turbulent advection.
\newblock {\em arXiv preprint nlin/0207058}, 2002.

\bibitem{gawedzki2000phase}
K.~Gaw\k{e}dzki and M.~Vergassola.
\newblock Phase transition in the passive scalar advection.
\newblock {\em Phys. D: Nonlin. Phen.}, 138:63--90, 2000.

\bibitem{chaves2003lagrangian}
M.~Chaves, K.~Gaw\k{e}dzki, P.~Horvai, A.~Kupiainen, and M.~Vergassola.
\newblock Lagrangian dispersion in {G}aussian self-similar velocity ensembles.
\newblock {\em J. Stat. Phys.}, 113(5):643--692, 2003.

\bibitem{doob1984classical}
J.~Doob.
\newblock {\em Classical potential theory and its probabilistic counterpart},
  volume 262.
\newblock Springer, 1984.

\bibitem{feller1991introduction}
W.~Feller.
\newblock {\em An introduction to probability theory and its applications},
  volume~2.
\newblock J. Wiley \& Sons, 1991.

\bibitem{gawedzki2008soluble}
K.~Gaw\k{e}dzki.
\newblock Soluble models of turbulent transport.
\newblock In {\em Non-equilibrium statistical mechanics and turbulence}, number
  355. 2008.

\bibitem{evans2012introduction}
L.~Evans.
\newblock {\em An introduction to stochastic differential equations},
  volume~82.
\newblock Am. Math. Soc., 2012.

\bibitem{pavliotis2014stochastic}
G.~Pavliotis.
\newblock {\em Stochastic processes and applications: diffusion processes, the
  {F}okker-{P}lanck and {L}angevin equations}, volume~60.
\newblock Springer, 2014.

\bibitem{kent1978some}
J.~Kent.
\newblock Some probabilistic properties of {B}essel functions.
\newblock {\em Ann. Probab.}, pages 760--770, 1978.

\bibitem{chevillard2015peinture}
L.~Chevillard.
\newblock {\em Une peinture al{\'e}atoire de la turbulence des fluides}.
\newblock {HDR} thesis, ENS Lyon, 2015.

\bibitem{pereira2016dissipative}
R.~Pereira, C.~Garban, and L.~Chevillard.
\newblock A dissipative random velocity field for fully developed fluid
  turbulence.
\newblock {\em J. Fluid Mech.}, 794:369--408, 2016.

\bibitem{chevillard2019skewed}
L.~Chevillard, C.~Garban, R.~Rhodes, and V.~Vargas.
\newblock On a skewed and multifractal unidimensional random field, as a
  probabilistic representation of {K}olmogorov’s views on turbulence.
\newblock In {\em Ann. H. Poincaré.}, volume~20, pages 3693--3741. Springer,
  2019.

\bibitem{reneuve2020flow}
J.~Reneuve and L.~Chevillard.
\newblock Flow of spatiotemporal turbulent-like random fields.
\newblock {\em Phys. Rev. Lett.}, 125(1):014502, 2020.

\bibitem{bezanson2017julia}
J.~Bezanson, A.~Edelman, S.~Karpinski, and V~Shah.
\newblock Julia: A fresh approach to numerical computing.
\newblock {\em SIAM review}, 59(1):65--98, 2017.

\bibitem{considera2023zenodo}
A.~Considera.
\newblock {M}ulti{F}ractal{F}ields.jl: v0.2.1.
  https://doi.org/10.5281/zenodo.8056736, June 2023.

\end{thebibliography}
\end{document}


\title{Supplemental material: Spontaneous stochasticity in the presence of intermittency}
\author{André Luís Peixoto Considera}
\affiliation{\impaaddress}
\affiliation{\ceaaddress}
\author{Simon Thalabard}
 \affiliation{\inphyniaddress}
\date{\today}

\maketitle

\section{Separation-times  in the 1D Kraichnan model ($\gamma=0$)}
\label{sec:sskraichnan}
This section recalls salient properties of spontaneous stochasticity in the Kraichnan model \cite{gawedzki2002soluble,gawedzki2000phase}, but explicitly addressed from the viewpoint of pair-separation times\cite{chaves2003lagrangian}, and \corr focused on the  unidimensional setting \rroc.
Kraichnan flows  correspond to the case of a vanishing intermittency parameter $\gamma =0$, hence prescribing  the velocity field as Gaussian  random flow. The dynamics  for $N$ tracers becomes (omitting $\gamma=0$ superscripts)
\be
	dX_{i} = u_{\eta}^{(\xi)}\left( X_i, dt \right) +\sqrt{ 2\eta}\; B_i(dt),
	\label{eq:dynamics_g}
\ee
with initial positions $X_i(0,x_i)=x_i$, and where $i=1..N$ denotes  tracer indexing.
As for the advecting Markovian field, it reduces to
\be
	u^{(\xi)}_\eta(x,dt)  = \int_{\mathbb R} L^{(\xi)}_{\eta}(x-y)\, W_1(dy,dt)
	\label{eq:MarkovGaussian},
\ee
It  is a centered homogenous Gaussian stochastic process with correlation function
\be
		 \av{u^{(\xi)}_\eta(t,x)u^{(\xi)}_\eta(t^\prime,x^\prime)} = \text{min}(t,t^\prime) \, C_\eta^{(\xi)}(x-x^\prime)
\ee
The spatial part  satisfies 
\be
  C_\eta^{(\xi)}(r)  
	 \sim 1-|r|^{\xi}  \text{ for $\eta \le |r| \le 1$},
\label{eq:correlation}
\ee
  and is also prescribed by
\be
	 C_\eta^{(\xi)}(r) = \int_{\mathbb R} L^{(\xi)}_{\eta}(r-y)L^{(\xi)}_{\eta}(-y) dy.
\ee
To quantify the spontaneous stochastic property 
\be
	\mathbb P \left[ \tau_1 < \infty \right] \to 1 \text{ as $\eta,\kappa \to 0$},
	\label{eq:ss}
\ee
we consider the separation 
$R(t,r_0) := X_2(t,r_0) -X_1(t, 0)$, and define the separation times  to some large scale $\lambda=O(1)$ as
\be
	\label{eq:exit_1d}
	\begin{split}
		&\tau_{\lambda}(\eta):= \inf\left \lbrace t: | R(t,\eta) | > \lambda \right\rbrace.
	\end{split}
\ee
The spontaneous stochastic condition \eqref{eq:ss} then becomes
\be
	\mathbb P \left[ \tau_\lambda < \infty \right] \to 1 \text{ as $\eta  \to 0$}.
	\label{eq:ss_1d}
\ee
\subsection{Separation-time statistics.}
From the  potential theory for Markov processes, the statistics of the random times \eqref{eq:exit_1d} \corr are linked  \rroc to solutions of  partial differential equations involving the backward operator  \cite{doob1984classical,feller1991introduction,gawedzki2008soluble,evans2012introduction}
\be
\label{eq:L22}
\begin{split}
	\mL_2 & = \left(1/2 +\eta \right)\left( \partial^2_{11} +  \partial^2_{22}\right) + C_\eta(x_1-x_2) \partial^2_{12}\\
	& =\left(\Done-\dfrac{\Dtwo}{2} \right) \partial^2_{xx} + 2 \Dtwo\, \partial^2_{rr},\\	
	 \text{with} &\hs \Dtwo(r,\eta):=  \dfrac{1}{2}|r|^\xi + \eta. \text{ for } \eta\le|r|\ll 1\\
	\text{ and }
	& \Done(\eta) :=  \dfrac{1}{2} + \eta.
\end{split}
\ee
In the previous equation, the second line uses the variables 
$r:= x_2-x_1,\, x:=(x_1+x_2)/2$, respectively representing the algebraic relative distance and the center of mass coordinates. The operator $\label{eq:L22}$ prescribes in particular the average of the process $Y(t):=y(X_1(t),X_2(t))$ where $y$ is smooth differentiable function $\mathbb R^2 \to \mathbb R$ as
\be
	\av{Y(t)} - Y(x_1,x_2) = \int_0^t dt\, \av{\mL_2 \;y (X_1,X_2)},
\ee
where $(X_1, X_2)$ satisfy 	\eqref{eq:dynamics_g}.

The average exit-time $y(r):=\av{\tau_\lambda(r)}$, \corr where  $\tau_\lambda(r)$ is defined by Eq.~\eqref{eq:exit_1d}\rroc, is obtained by solving the boundary-value problem (BVP)
\be
	 \mathcal L_2 y(r) = -1 \text{ subject to   } y(\lambda)=0 \text{ and } y(0)=0.
	\label{eq:pb1}
\ee

The  thermal diffusion terms  controls the separation statistics on scale $r<\eta$: The average exit time from $0$ to $\eta$ is 
$t_\eta \sim \eta \to 0$. As such,  the BVP \eqref{eq:pb1} becomes formally equivalent, in the limit of vanishing  $\eta$, to the simpler BVP \cite{gawedzki2008soluble,chaves2003lagrangian}
\be
	 \mathcal L_2 y(r) = -1 \text{ subject to   } y(\lambda)=0 \text{ and } y'(\eta)=0,
\ee
The boundary condition $y'(\eta)=0$ comes from the  prescription that trajectories reflect from each other upon collision \cite{evans2012introduction,pavliotis2014stochastic}.
For positive argument $r>\eta$,  one expects the thermal noise contribution to become negligible. Further assuming that the scaling \eqref{eq:correlation}
 exactly holds for $\eta<r<\lambda$ \cite{gawedzki2008soluble},   the change of variable $ u = r^{1-\xi/2}$ transforms the previous equation into  one involving the  heat kernel  in effective dimension  $d_e = (2-2\xi)/(2-\xi)$ and effective diffusivity $D_\xi=(1-\xi/2)^2$ as
\be
\label{eq:av}
\begin{split}
&	D_\xi \left(u^{1-d_e} \partial_u u^{d_e-1} \partial_u\right) y=-1\\
& \text{ subject to   } y(\lambda)=0 \text{ and } y'(\eta) = 0.
\end{split}
\ee
 Eq.~\eqref{eq:av} then yields 
\be
	\left\langle \tau_\lambda(\eta)\right\rangle \underset{\eta\to0}{\sim}
	\begin{cases}
	&  \dfrac{\lambda^2}{2d_eD_\xi} =O(1)\text{ if $d_e> 0$}\\
	& \dfrac{\lambda^2}{ d_e(2-d_e)D_\xi}  \left(\dfrac{\eta}{\lambda}\right)^{d} \to \infty \text{ if $d_e > 0$ }\\
	& \dfrac{\lambda^2}{ 2 D_\xi}  \log \dfrac{\lambda}{\eta} \to \infty \text{ if $d_e = 0$ }
	\end{cases}.
	\label{eq:tauav}
\ee
The finiteness of the average entails the spontaneously stochastic property \eqref{eq:ss}:
The trajectories are spontaneously stochastic if and only if $d_e>0$ corresponding to  $\xi <1$.\\

\subsection{Beyond averages.}
In the spontaneously stochatic regime $\xi<1$, the exit-time statistics are fully prescribed by the 
boundary value problem
\be
\begin{split}
	&  L_2 y(u) -s y =0  \\
	&\text{subject to }  y(\lambda)=0 \text{ and } u^{d_e-1}y'|_{u=0}=0,
	\label{BVPsturm}
\end{split}
\ee
which yields the characteristic function
\be
	\chi(\eta,s)=y(u_0 ) = \left\langle e^{-s\tau_\lambda(\eta)}\right\rangle \text{ with } \eta = u_0^{2/(2-\xi)}.
\ee
The  BVP  ~\eqref{BVPsturm} has an explicit solution in terms of Bessel function \cite{kent1978some}:
\be
	 \chi(\eta,s) = \left(\dfrac{\lambda}{\eta}\right)^{\nu} \dfrac{I_\nu(\eta x)}{I_\nu(\lambda x)} \underset{\eta\to 0} \sim   \dfrac{(\lambda x)^\nu}{2^\nu\Gamma(\nu+1) I_\nu(\lambda x)}, 
\ee
with the shorthands $x:= (s/D_\xi)^{1/2}$,  $\nu:=d_e/2-1$, and $I_\nu$  the modified Bessel function of the first kind. %
The characteristic function  prescribes all the  moments of the exit times.  For example, small-$s$ expansion of the last rhs term yields
\be
\nonumber
	 \chi(0 ,s) = 1 - s \underbrace{\dfrac{\lambda^2}{4 (\nu +1) D_\xi}}_{\left\langle \tau_\lambda(0)\right\rangle}+\dfrac{s^2}{2}\underbrace{ \dfrac{\lambda^4(\nu+3)}{2^4 (\nu+2)(\nu +1)^2 D_\xi^2}}_{\left\langle \left(\tau_\lambda(0)\right)^2\right\rangle} + \cdots,
\ee
from which one retrieves Eq.~\eqref{eq:tauav},  further gets 
\be 
	\text{Var}(\tau_\lambda(0)) = \dfrac{\lambda^4}{16 (\nu +1)^2 (\nu+2) D_\xi^2}.
 \ee 
and deduce   from  Tauberian theorems \cite{feller1991introduction} the asymptotic exponential behavior
\be
	\mathbb P(\tau_\lambda(0) \in d\tau) \underset{\tau \to \infty}\sim \dfrac{ d\tau\;e^{-\tau /\av{\tau_\lambda(0)}}}{\av{\tau_\lambda(0)}}.
\ee

\section{$\gamma$-BVP and mean-field Ansatz}
\label{eq:effoperator}
We here provide technical details on the derivation of the driving Hurst parameter \eqref{eq:xigamma} under the mean-field Ansatz.
\corr
For a prescribed environment $Y$,
and similar to the Gaussian case, the formal use of  It\^o calculus   in the SDE representation  prescribes the two-particle generator
\be
\begin{split}
& \mathcal L_{2}^Y(x_1,x_2)= \dfrac{e^{2\gamma Y(x_1)}}{2Z^{2}} \left[\partial_{x_{1}x_{1}}+2(1-r^{\xi})\partial_{x_{1}x_{2}}+\right.\\
&+ \left. \left((1-r^{\xi})^{2}+(1-(1-r^{\xi})^{2})e^{2\gamma\Delta Y  )}\right)\partial_{x_{2}x_{2}} \right],
\end{split}
\ee
with the shorthand $\Delta Y=Y(x_2)-Y(x_1)$.
Assuming that the average separation time $T_1^Y$ depends on the separation $r=x_2-x_1$ only,  then it  solves  the boundary-value problem
\be
\label{eq:pb2_aux}
	\begin{split}
	  & \mathcal L_{2}^Y T_1^Y = -1 \\
	& \text{ subject to } \partial_r T_1^Y(r=\eta) =0, \;\;\; T_1^Y(r=1) =0 
	\end{split}
\ee
\rroc
for the  GMC-realization-dependent operator 
\be
\nonumber
\begin{split}
	& \mathcal L_{2}^Y(x_1,r): =\dfrac{e^{2\gamma Y(x_1)}}{2Z^2} \left(r^\xi+ e^{2\gamma \Delta Y} \left(2-r^{\xi}\right)  \right)r^{\xi} \partial_{rr}.
\end{split}
\ee
Eq~.\eqref{eq:pb2_aux} is the analog of Eq.~\eqref{eq:pb1} obtained in the Gaussian case.
From this  definition, we recast Eq.~\eqref{eq:pb2_aux} as
\be
	\left(r^\xi+ e^{2\gamma \Delta Y} \left(1-\dfrac{r^{\xi}}{2}\right)  \right)r^{\xi} \partial_{rr} T_1^Y = - Z^2e^{-2\gamma Y(X_1)},
\ee
and, averaging over the random GMC,  get
\be
	\label{eq:av_aux}
	\avY{\left(r^\xi+ e^{2\gamma \Delta Y} \left(2-r^{\xi}\right)  \right)r^{\xi} \partial_{rr} T_1^Y}= -2 Z^4.
\ee
Here, the rhs stems  from  the observation 
\be
\avY{e^{-2\gamma Y}} = \avY{e^{2\gamma Y(X_1)}}= Z^2
\ee
Under the mean-field Ansatz
\be
	\label{eq:meanfield}
	\avY{e^{\gamma \Delta Y} T_1^Y}= \avY{e^{\gamma \Delta Y}} \tau_1 
\ee
involving $\tau_1 = \avY{T_1^Y}$,
 Eq.~\eqref{eq:av_aux} formally becomes
\be
	\label{eq:av_aux2}
	\left(r^\xi+ \avY{e^{2\gamma \Delta Y}} \left(2-r^{\xi}\right)  \right)r^{\xi} \partial_{rr} \tau_1 = -2 Z^4.
\ee
Now, we compute 
\be
	\label{eq:av_aux3}
	\begin{split}
		\avY{e^{2\gamma \Delta Y}} & = e^{4\gamma^2 \avY{\Delta Y^2}}\\
		 & = Z^4 e^{-4\gamma^2 \avY{Y(X+r/2)Y(X-r/2)}}\\
		 & = Z^4 r^{4 \gamma^2},
	\end{split}
\ee
where the last identity stems \corr from \rroc the definition of  $Y$ as a monofractal Gaussian field with vanishing Hurst parameter,
with correlation function $C^{(0)}(r) =\log(1/r) \mathbf 1_{r<1}$.
Using the estimate \eqref{eq:av_aux3} in Eq.~\eqref{eq:av_aux2}, and collecting the terms with
order  $$Z^4 = O(\eta^{-4\gamma^2})  \underset{\eta \to 0}\to\infty$$   yields 
\be
\begin{split}
	&  \mathcal L^*_2 \tau_1 = -1 \;\;\; \text{   with   } \tau_1(1)=0 \text{ and } \partial_r \tau_1(0)=0,\\
	& \text{ for } 	\mathcal L_{2}^*(r): = \left(1-\dfrac{r^\xi}{2}\right) r^{\xi+4\gamma^2} \partial_{rr}.
\end{split}
\label{eq:op3},
\ee
 together with the effective driving Hurst parameter 
\be
	\label{eq:xigamma}
	\xi_{\gamma} := \xi + 4 \gamma^2 > \xi.  
\ee

\section{Field-based numerics}
\label{sec:numerics}
Following previous strategies \cite{chevillard2015peinture,pereira2016dissipative,chevillard2019skewed,reneuve2020flow}, our numerical implementation of the multifractal random flows
\be
	\begin{split}
	& u^{\xi,\gamma}_\eta(x,dt)  = \dfrac{1}{Z}\int_{\mathbb R} L^{(\xi)}_{\eta}(x-y)\,e^{\gamma Y(y)}  W_1(dy,dt),\\
	& \text{for}\;
	Y(y) := \int_{\mathbb R} L_{\eta}^{(0)}(y-z) W_2(dz),\;\; Z:= e^{ \gamma^2 \mathbb E\left({Y^2}\right)}
	\end{split}
	\label{eq:MarkovGMC}
\ee
uses the periodization of the space domain, in order to take advantage of  Fast-Fourier transforms to compute the convolution products. 
The regularization scale $\eta$ is  prescribed by  the numerical resolution of the field,  that is $\eta = 4\pi/N$.
Compared to the previous numerical implementions \cite{chevillard2015peinture, reneuve2020flow} the one specific feature of our numerics is the choice of the  kernel $L^{(\xi)}$, which is defined through a Fourier transform and does not have simple expression in physical space. In this work, \corr we rather prefer to  explicitly  specify the correlation function as in Eq.~\eqref{eq:correlation},  as it is the correlation function that  controls  the spontaneously stochastic nature of  Gaussian Kraichnan flows. \rroc

Our $2\pi$-periodic version  of the field ~\eqref{eq:MarkovGMC}  is then
\be
	\begin{split}
	& u_{\eta}^{\xi,\gamma}(x,dt) :=   \sum_{|k| < 4\pi/\eta }e^{ikx} L^{(\xi)}(k) M^{(\gamma)}_\eta(k,dt) , \\
	& \text{with }L^{(\xi)}(k):= \left| \dfrac{2}{\pi}\int_{0}^1 \; dr \;C_\xi(r) \cos (kr) \right|^{1/2},
	\end{split}
	\label{eq:torusgmc}
\ee
featuring the Fourier representation of the GMC kernel
\begin{equation}
	\begin{split}
	  M^{(\gamma)}_{\eta}(k,dt) &:= \dfrac{1}{2\pi Z_\eta}\int_{-\pi}^\pi \; e^{-ikx}  e^{\gamma Y} \; W_1(dx,dt),\\
	&  Z:=e^{  \gamma^2 \avY{Y^2}}.
	\end{split}
\end{equation}
The latter formula involves the two random processes
\be
\begin{split}
	&W_1(x,dt):=  \sum_{|k| < 4\pi/\eta} e^{ikx} \hat W^{(1)}_k(dt),\\
\text{and} \;	 &Y:=  \sum_{|k| < 4\pi/\eta} e^{ikx} L^{(0)}(k) \hat W^{(2)}_k(dt) 
\end{split}
\label{eq:W1}
\ee
featuring  two mutually independent complex white noises  $\hat W^{(1,2)}_k(dt)$. 
\corr
Specifically, each 
$ \hat W^{(i)}_k(t) $  are  complex centered Gaussian processes, mutually independent except for the Hermitian symmetry
$\hat W^{(i)}_{-k}(t) = \hat W^{(i),*}_{k}(t)$. They have the  explicit correlation 
 \be \mathbb E \left(\hat W^{(i)}_k(t) \hat W^{(j)}_{-k}(t)\right) = \delta_{ij}\delta_{kk'} \min(t,t').
\ee
\rroc
The normalizing factor is then simply estimated as 
\be
	 Z(\eta) = \sum_{|k| < 4\pi/\eta} |L^{(0)}(k)|^2 =O( \eta^{-\gamma^2}). 
\ee
\color{black}
All the numerics were performed using the Julia language \cite{bezanson2017julia}. The figures presented in this  manuscript can be replicated using the package MultiFractalFields.jl. \cite{considera2023zenodo}
\color{black}

\bibliographystyle{unsrt}
\bibliography{Biblio}